\documentclass[11pt]{article}
\usepackage[authoryear,round,sort]{natbib}
\usepackage{sublabel}
\usepackage{latexsym}
\usepackage{amsthm}
\usepackage{graphicx}
\usepackage{amsmath}
\usepackage{pstricks}
\usepackage{fullpage}
\newpsobject{showgrid}{psgrid}{subgriddiv=1,griddots=10,gridlabels=6pt}

\newcommand {\norm} [1] { \lVert #1 \rVert}

\newcommand {\abs} [1] {\left| #1 \right|}

\numberwithin{equation}{section}

\title{Recurrent motions within plane Couette turbulence}
\date{December 2006}
\author{D. Viswanath \thanks{Department of Mathematics,
University of Michigan, 
530 Church Street, Ann Arbor, MI 48109, U.S.A. }}

\begin{document}
\maketitle

\begin{abstract}
 The phenomenon of bursting, in which streaks in turbulent boundary
 layers oscillate and then eject low speed fluid away from
 the wall, has been studied experimentally, theoretically, and
 computationally for more than 50 years because of its importance to
 the three-dimensional structure of turbulent boundary layers. We
 produce five new three-dimensional solutions of turbulent plane
 Couette flow, one of which is periodic while four others are relative
 periodic. Each of these five solutions demonstrates the break-up and
 re-formation of near-wall coherent structures. 
 Four of our solutions are periodic but with drifts in the
 streamwise direction. More surprisingly, two of our solutions are
 periodic but with drifts in the spanwise direction, a possibility
 that does not seem to have been considered in the literature. We
 argue that a considerable part of the streakiness observed
 experimentally in the near-wall region could be due to spanwise
 drifts that accompany the break-up and re-formation of coherent
 structures. We also compute a new periodic solution of plane Couette
 flow that could be related to transition to turbulence.

 The violent nature of the bursting phenomenon implies the need for
 good resolution in the computation of periodic and relative periodic solutions
 within turbulent shear flows. We address this computationally
 demanding requirement with a new algorithm for computing relative
 periodic solutions one of whose features is a combination of two
 well-known ideas --- namely the Newton-Krylov iteration and the
 locally constrained optimal hook step. Each of our six solutions
 is accompanied by an error estimate.

 In the concluding discussion, we discuss dynamical principles that
 suggest that the bursting phenomenon, and more generally fluid
 turbulence, can be understood in terms of periodic and relative
 periodic solutions of the Navier-Stokes equation.
\end{abstract}

\newpage

\section{Introduction}
Turbulent boundary layers are characterized by a viscous sublayer,
in which the mean streamwise velocity  nearly equals the
distance from the wall in wall units, a buffer layer, and a
logarithmic boundary layer. The impressive agreement of theoretical
predictions of the mean streamwise velocity in the viscous sublayer
and the logarithmic boundary layer with experiment \citep[p. 273]{MY}
and with computation \citep{KMM} can be considered an outstanding
success in the effort to understand turbulence in fluid flows.  It was
initially believed that the flow in the viscous sublayer was laminar,
but experiments in the 50s and 60s lead to the conclusion that random
fluctuations exist in this layer even though the mean streamwise
velocity in this layer had a laminar profile \citep[p. 270]{MY}.

The phenomenon of bursting became evident during investigations of the
viscous sublayer and the buffer region \citep{KRSR, KTS}. In this
phenomenon, streaks in the near-wall region break-up and re-form in a
quite striking manner. Much of the turbulent energy production occurs
in the buffer and viscous layers \citep{KRSR} and the
three-dimensional structure of turbulent boundary layers appears to be
intimately related to bursting. Because of this connection and because
of the striking nature of the phenomenon itself, bursting has been the
subject of numerous experimental \citep{AS, BTAA, KRSR, KTS, SM} and
computational or theoretical studies \citep{HKW, HLB, IT, ITx, JKSNS, KK,
SH}. There has been some discussion in the literature of exactly what
is meant by bursting \citep{IT, JKSNS}. In this paper, bursting will
always refer to the break-up and re-formation of coherent structures,
such as streaks, in turbulent buffer regions.

Although bursting has been much studied, its dynamics has proved
elusive. A large number of {\it mechanisms} have been proposed
to explain bursting \citep{HKW, IT, KRSR, KTS, SH}. The word
mechanism in this context does not refer to new physical principles.
There is no doubt at all that the incompressible Navier-Stokes 
equation is adequate to explain the dynamics of bursting, and
the physics of bursting is the physics that goes into that 
equation and its boundary conditions. However, as is well known, the
nature of the solutions of the Navier-Stokes equation in the turbulent
regime is poorly understood. These mechanisms try to 
provide a way to {\it approximately} understand some solutions of
the Navier-Stokes equation.

Although approximate, some mechanisms can be useful for computing
exact solutions as shown by the construction of traveling wave and
steady solutions of channel flows by Waleffe \citep{Waleffe3,
Waleffe4, Waleffe5}. The method introduced in that body of work was later
adapted to pipe flows \citep{FE, WK}. However, the break-up and
re-formation of coherent structures cannot be studied using traveling
waves or steady solutions.

It is certainly very desirable
to find exact solutions of the Navier-Stokes equation that correspond
to bursting. These solutions will provide a  solid and reliable
route to understanding the dynamics of bursting. In this context, it
is noteworthy that the self-sustaining process mechanism indeed
suggested the existence of periodic solutions that correspond to
bursting \citep{HKW, KK, Waleffe2}. We follow
\citet{HKW} and conduct our computations using plane Couette flow
at a Reynolds number ($Re$) of 400. In plane Couette flow two parallel
walls move in opposite directions with equal speed and drive the fluid
in-between.  The Reynolds number is based on half the separation
between the walls and half the difference between the wall
velocities. To render the computational domain finite, we assume the
domain to be periodic in the streamwise and spanwise directions, with
periods equal to $2\pi\Lambda_x$ and $2\pi\Lambda_z$. Unless otherwise
stated, our computations use $\Lambda_x = 0.875$ and $\Lambda_z = 0.6$
to facilitate comparison with earlier computations and because of
particular advantages of this box described in \citep{HKW}.  The walls
are assumed to be at $y = \pm 1$, with the upper wall moving in the
$x$ or streamwise direction in the positive sense with speed equal to
$1$.

It is known experimentally that the details of bursting in the near-wall
region are remarkably similar over a very wide range of Reynolds numbers
\citep{SM}.
Thus our use of $Re=400$ in plane Couette flow is an acceptable choice.
Turbulent spots have been observed in plane Couette flow experiments at 
$Re=360$ \citep{BTAA}.

\begin{table}[tb]
\begin{center}
\begin{tabular}{c|c|c|c|c|c|c|c|c}
Marker & Label &$s_x/2\pi\Lambda_x$ & $s_z/2\pi\Lambda_z$ & $T/T^+$
& $\lambda_{max}$ & $Re_\tau$ & $(2L, M, 2N)$ & error\\ \hline
$\ast$ & $P_1$ & $0$ & $0$ & $82.7$/$167$ & 
$24.5$ & $56.9$ & $(32,64,48)$ & $10^{-7}$\\ \hline
o & $P_2$ & $.15$ & $-.28$ & $70.9$/$197$ & 
$12.2$ & $66.3$ & $(48,64,48)$ & $10^{-5}$\\ \hline
$\Box$ & $P_3$ & $.03$ & $0$ & $77.1$/$216$ & 
$4.4+\iota 3.6$ & $66.9$ & $(64,56,48)$ & $10^{-5}$\\ \hline
$\Diamond$ & $P_4$ & $.01$ & $.08$ & $102.7$/$287$ & 
$5.1+\iota 16.4$ & $66.8$ & $(48,48,32)$ & $10^{-3}$\\ \hline
$\triangle$ & $P_5$ & $.22$ & $0$ & $91$/$266$ & 
$-11.5$ & $68.4$ & $(48,64,48)$ & $10^{-4}$\\ \hline
$\nabla$ & $P_6$ & $0$ & $0$ & $87.9$/$259$ & 
$-14.6$ & $68.6$ & $(48,64,48)$ & $10^{-5}$\\ \hline
\end{tabular}
\end{center}
\caption[xyz]{This table gives data for six periodic and relative periodic
solutions of plane Couette flow labeled $P_1$ through $P_6$. The
markers in the first column are used to distinguish between these
solutions in later plots. The shifts $s_x$ and $s_z$ are explained in
the text.  The $T/T^+$ column gives the period, with $T^+$ being the
period in wall units. The $\lambda_{max}$ column gives the maximum
characteristic multiplier. The last three columns report the
frictional Reynolds number, the resolution of the computational grid, and
the relative error in the computation.}
\label{table-1}
\end{table}

Table \ref{table-1} reports data for six periodic or relative periodic
solutions that we computed. The final velocity field obtained by
integrating the initial velocity field over one full period equals the
initial velocity field for periodic solutions. But for relative
periodic solutions the final velocity field equals the initial
velocity field after shifts equal to $s_x$ and $s_z$ in the streamwise
and spanwise directions, respectively. These shifts, which are
reported in Table \ref{table-1}, are both $0$ for $P_1$ and
$P_6$. Thus both those solutions are periodic. The equations of plane
Couette flow are unchanged by translations in the streamwise and
spanwise directions. The existence of relative periodic solutions is
possible because of those invariances.

Frictional velocity and frictional length can be obtained using the
mean shear at the wall \citep[p. 265]{MY}.  Those quantities are the
basis of wall units.  Throughout this paper, wherever wall units are
used, the mean shear is obtained by averaging at the upper wall over
one single period of a periodic or relative periodic solution.
Following standard practice, the use of wall units is signaled by
using $+$ as a superscript. The frictional Reynolds number $Re_\tau$
was obtained using the mean shear at the upper wall and the distance
between the two walls. Table \ref{table-1} shows that the $Re_\tau$
for $P_1$ is significantly lower than it is for the other
solutions. This is because $P_1$ alone is related to transition to
turbulence, while all the others are related to bursting.

\cite{KK} found a periodic solution of plane Couette
flow with period equal to $85.5$ which appears similar to
$P_1$. However, their solution satisfies the shift-reflection and
shift-rotation symmetries of plane Couette flow
\citep{Kawahara}. Although $P_1$ has zero mass flux in
the streamwise direction like their solution, it is very far from
satisfying either symmetry as will be shown. Therefore $P_1$ is a new
solution. The bursting solutions $P_2$ through $P_6$ are all new, and
we will presently argue that these are the first computations of
bursting periodic or relative periodic solutions that demonstrably
correspond to solutions of the Navier-Stokes equation.

All the characteristic multipliers $\lambda_{max}$ listed in Table
\ref{table-1} are outside the unit circle, which implies instability of
the solutions $P_1$ through $P_6$. Yet we report relative errors for
all these solutions.  A relative error of $10^{-7}$ implies that the
computed solution matches a solution of the Navier-Stokes equation up
to at least $7$ digits.  The manner in which these error estimates were
found is explained in Section 3. Significantly, these error estimates can
be verified using any good DNS (direct numerical simulation) code in
spite of the instability of the underlying solutions. Such
quantitative reproducibility is a step forward for turbulence
computations.

The numbers $2L$, $M$, and $2N$ in Table \ref{table-1} give the number
of grid points in the $x$, $y$, and $z$ directions. We used a Fourier
grid in the $x$ and $z$ directions and a Chebyshev grid in the $y$
direction. Good spatial resolution is the key to finding solutions
that are not numerical artifacts. \cite{HKW} and \cite{KK} used $(2L,M,2N) = (16,32,16)$
in their plane Couette flow computations.  The evidence for striking
recurrences and the existence of periodic solutions offered in these
works is significant. However, estimates for spatial discretization
error described in Section 3 imply that the spatial discretization
error with $(2L,M,2N) = (16, 32, 16)$ for bursting solutions is at
least $5\%$ and can be twice as much. 

A section 2, we describe a new method for finding periodic and
relative periodic solutions. The number of degrees of freedom that
determine the initial velocity field for $P_3$ is $319790$ and the
number of degrees of freedom for $P_2$, $P_5$, and $P_6$ is
$273918$. These numbers exceed the number of degrees of freedom in any
earlier computation of periodic solutions by at least a factor of
$20$, and our method can also compute relative periodic solutions. One
feature of the method is a combination of the Newton-Krylov iteration
\citep{Kelley, SNGS} with the locally constrained optimal hook step
\citep{DS}. The locally constrained optimal hook step is related to
the Levenberg-Marquardt procedure and is a well-established idea in
optimization. However, its possibilities seem to have been largely
overlooked in computations of periodic and steady solutions. The
combination of Newton-Krylov iterations with the locally constrained
optimal hook step is simple, but powerful. It is much more effective
than the often used damped Newton iteration.

In Section 4, we develop the connection of the computations summarized
in Table \ref{table-1} to the dynamics of the bursting phenomenon.
There has been some discussion about whether the near-wall bursting
observed experimentally is due to the advection of coherent objects or
to the break-up and re-formation of coherent objects
\citep{JKSNS}. The relative periodic solutions reported in Table
\ref{table-1}, and especially the spanwise drifts of $P_2$ and $P_4$,
will be shown to be significant in this respect.

In the concluding Section 5, we discuss our belief that a good
route to understanding the dynamics of a differential equation is by
computing its solutions and recognizing the relationship between those
solutions. We discuss dynamical principles that suggest that
infinitely many periodic and relative periodic motions can be found
within turbulent flows. About half a century ago, a common belief was
that linearly stable solutions are observed in nature and in
experiment, while the unstable ones are not. Although all these
infinitely many periodic and relative periodic motions are certain to be
linearly unstable, their instability is only a manifestation of the
instability of turbulent flows. In spite of their instability, these
solutions are relevant both to natural phenomena and experiment as
we demonstrate in Section 4 and as we argue in Section 5.

The idea of understanding phenomena using well-resolved and linearly
unstable nonlinear solutions is beginning to take root in recent
research on the transition to turbulence in shear flows
\citep{Kerswell}. Linearly unstable nonlinear traveling waves
have been observed in a pipe flow experiment \citep{HD}. Although
transition to turbulence in pipe and channel flows is still an
unsolved problem, there can be little doubt that the nonlinear traveling
waves such as the ones observed by \cite{HD} will be an important
part of an eventual solution of the transition problem. Periodic and
relative periodic solutions are the next step for understanding
turbulent phenomena such as bursting.

\section{A numerical method for finding relative periodic solutions}
This section describes a numerical method for finding periodic and
relative periodic solutions in plane Couette flow.  Although the
description of the numerical method is specific to plane Couette flow,
it can be easily adapted to other partial differential equations.  Our
method has three new aspects. Firstly, we describe a way to find good
initial guesses. Secondly, we show how to set up the Newton equations
for finding relative periodic solutions. Thirdly, we show how to
modify the Newton-Krylov procedure to compute the locally 
constrained optimal hook step.

The Navier-Stokes equation for incompressible flow takes the form
\begin{equation}
\partial {\bf u}/\partial t + ({\bf u}.\nabla){\bf u} =
-(1/\rho)\nabla p + (1/Re)\triangle {\bf u}
\label{eqn-2-1}
\end{equation}
 with the
incompressibility constraint implying $\nabla.{\bf u} = 0$.  For plane
Couette flow, the boundary conditions are ${\bf u} = (\pm 1, 0, 0)$ at
the walls (which are at $y=\pm 1$) in the $y$ or wall-normal
direction, and periodic in the other two directions with periods equal
to $2\pi\Lambda_x$ and $2\pi\Lambda_z$.  The equation cannot be viewed
as a dynamical system when written in this form because the velocity
field ${\bf u}$ has to satisfy the zero divergence condition and there
is no explicit equation for evolving the pressure $p$ in time. It is
necessary to rewrite the equation as a dynamical system for the
purpose of computing periodic and relative periodic solutions.

The pressure term can be completely eliminated by recasting 
\eqref{eqn-2-1} in terms of  the wall-normal velocity $v(x,y,z)$,
the wall-normal vorticity $\eta(x,y,z)$, and the mean components
$\bar{u}(y)$ and $\bar{w}(y)$ of the streamwise and spanwise
velocities. The boundary conditions become $\bar{u}(\pm 1) = \pm 1$,
$\bar{w}(\pm 1) = 0$, and $v(x, \pm 1, z) = v_y(x, \pm 1, z) = \eta(x,
\pm 1, z) = 0$. The velocity field ${\bf u}$ can be constructed from
$\bar{u}$, $\bar{w}$, $v$, and $\eta$ using $\nabla.{\bf u} = 0$.
The velocity component $v$ and the vorticity component $\eta$ are discretized
in the $x$ and $z$ directions using Fourier modes. For example,
\begin{equation}
v(x,y,z) = \sum_{\substack{-L< l < L\\-N<n<N}} \hat{v}_{l,n}(y)
\exp\Bigl(\frac{\iota l x}{\Lambda_x} + \frac{\iota n z}{\Lambda_z}\Bigr), 
\label{eqn-2-2}
\end{equation}
where the dependence on $t$ is not shown explicitly.
The Fourier modes $\hat{\eta}_{l,n}$ are defined by replacing $v$ by
$\eta$ in \eqref{eqn-2-2}. 

\cite{KMM} proposed a very good numerical
method for integrating the Navier-Stokes equation using this
formulation and we follow their approach. The Navier-Stokes equation
\eqref{eqn-2-1} becomes a dynamical system in this formulation.

We use a Fourier grid with $2L$ and $2N$ points in the $x$ and $z$
directions, but we always set the modes with $l=L$ or $n=N$ equal to
$0$. We use $M+1$ Chebyshev points in the $y$ direction. After spatial
discretization, the dynamical system has approximately $8LMN$ degrees
of a freedom. A count that takes into account the boundary conditions
on $\bar{u}$, $\bar{w}$, $v$ and $\eta$, the treatment of modes with
$l=L$ or $n=N$, and the fact that the mean components of $v$ and
$\eta$ are always zero gives $2(M-1) + (2M-4)((2N-1)(2L-1)-1)$ as the
exact number of degrees of freedom. It is important to get the number
of degrees of freedom exactly right. The spatially discretized system
can be thought of as a dynamical system of the form $\dot{X} = f(X)$,
where the $2(M-1) + (2M-4)((2N-1)(2L-1)-1)$ components of $X$ encode
$\bar{u}$, $\bar{w}$, $v$ and $\eta$. All the components of $X$ are
real numbers.

 Our code implements the nonlinear terms in advection, rotation, and
skew-symmetric forms. The viscous terms are treated implicitly and
the advection terms are treated explicitly when discretizing time.
The code employs dealiasing using  the $3/2$ rule in the $x$ and $z$
directions.
The code was tested using three-dimensional modes of the Orr-Sommerfeld
equation and in various other ways.

\begin{center}
\textit{2.1 Finding initial guesses}
\end{center}
\begin{figure}
\begin{center}
\includegraphics[height=2in,width=2in]{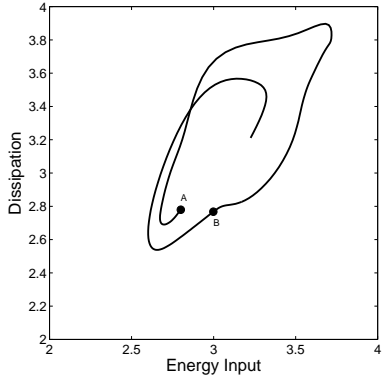}
\end{center}
\caption[xyz]{This figure shows the projection of a trajectory of plane
Couette flow onto the energy dissipation ($D$) and energy input($I$) plane}
\label{fig-1}
\end{figure}

To find an initial guess for a relative periodic solution, we begin
by looking at projections of trajectories to the energy dissipation
and energy input plane.
The rate of energy dissipation per unit volume for plane
Couette flow is given by
\begin{equation}
D = \frac{1}{8\pi^2 \Lambda_x\Lambda_z}\int_0^{2\pi\Lambda_z}
\int_{-1}^{+1} \int_0^{2\pi\Lambda_x} \abs{\nabla u}^2
+\abs{\nabla v}^2 + \abs{\nabla w}^2 \: dx\, dy\, dz
\label{eqn-2-3}
\end{equation}
and the rate of energy input per unit volume is given by
\begin{equation}
I = \frac{1}{8\pi^2 \Lambda_x\Lambda_z}\int_0^{2\pi\Lambda_x}
\int_0^{2\pi\Lambda_z} \frac{\partial u}{\partial y} \Bigl\lvert_{y=1}
+ \frac{\partial u}{\partial y} \Bigl\lvert_{y=-1} \: dx\, dz.
\label{eqn-2-4}
\end{equation}
For the laminar solution $(u,v,w) = (y,0,0)$, both $D$ and $I$ are
normalized to evaluate to $1$.

The trajectory that starts at $A$ in Figure \ref{fig-1} appears to
come close at $B$. But that close approach is not significant because
even distant points in phase space can coincide in a two-dimensional
projection. What we look for is the resemblance of the shape of the
trajectory from $B$ onwards with the shape of the trajectory from
$A$ onwards. In Figure \ref{fig-1}, the trajectory that begins at $A$
develops a type of protrusion in the upper right part of the figure,
but the trajectory that begins at $B$ does not. Therefore, $A$ would
not be a good initial guess for a relative periodic solution. If it 
were, we would examine velocity fields on the trajectory near $B$, and shift
them in the streamwise and spanwise directions to bring them as close
to $A$ as possible. The initial guesses for the period and the shifts
would be determined following that examination. 

\begin{center}
\textit{2.2 Newton equations for finding relative periodic solutions}
\end{center}
Given the Fourier representation \eqref{eqn-2-2} of $v(x,y,z)$, the
Fourier representation of $v(x+s_x, y, z+s_z)$ is given by
\begin{equation}
v(x+s_x, y, z+s_z) = \sum_{\substack{-L< l < L\\-N<n<N}}
\exp\Bigl(\frac{\iota l s_x}{\Lambda_x}\Bigr)
\exp\Bigl(\frac{\iota n s_z}{\Lambda_z}\Bigr)
\hat{v}_{l,n}(y) 
\exp\Bigl(\frac{\iota l x}{\Lambda_x} + \frac{\iota n z}{\Lambda_z}\Bigr). 
\label{eqn-2-5}
\end{equation}
Define the operators $\mathcal{T}_1$ and $\mathcal{T}_2$ by
\begin{align}
\sublabon{equation}
\mathcal{T}_1 v(x, y, z) = \sum_{\substack{-L< l < L\\-N<n<N}}
\frac{\iota l}{\Lambda_x} \hat{v}_{l,n}(y)
\exp\Bigl(\frac{\iota l x}{\Lambda_x} + \frac{\iota n z}{\Lambda_z}\Bigr)
\label{eqn-2-6a}\\
\mathcal{T}_2 v(x, y, z) = \sum_{\substack{-L< l < L\\-N<n<N}}
\frac{\iota n}{\Lambda_z} \hat{v}_{l,n}(y)
\exp\Bigl(\frac{\iota l x}{\Lambda_x} + \frac{\iota n z}{\Lambda_z}\Bigr).
\label{eqn-2-6b}
\end{align}
\sublaboff{equation}
The operators $\mathcal{T}_1$ and $\mathcal{T}_2$ are infinitesimal
generators of the group of translations along the streamwise and
spanwise directions.
The shift in \eqref{eqn-2-5} is given by $e^{s_x\mathcal{T}_1}e^{s_z\mathcal{T}_2} v(x,y,z)$.
To shift a velocity field given by $\bar{u}$, $\bar{w}$, $v$, and $\eta$
by $s_x$ along $x$ and by $s_z$ along $z$, we need only apply 
$e^{s_x\mathcal{T}_1}e^{s_z\mathcal{T}_2}$ to $v$ and $\eta$. To shift a velocity field
encoded by $X_0$, we can convert $X_0$ to $\bar{u}$, $\bar{w}$,
$v$, $\eta$, shift $v$ and $\eta$, and then convert back.
The shift operation on $X_0$ will be denoted by
$e^{s_x\mathcal{T}_1} e^{s_z\mathcal{T}_2} X_0$.

The Navier-Stokes equations for plane Couette flow are unchanged by
shifts in the $x$ and $z$ directions. In terms of 
$\dot{X} = f(X)$, this property becomes $f(e^{s_x\mathcal{T}_1}e^{s_z\mathcal{T}_2}X)
= e^{s_x\mathcal{T}_1}e^{s_z\mathcal{T}_2}f(X)$.

To find a relative periodic solution of plane Couette flow, we
have to find an initial velocity field $X_0$, a period $T$,
and shifts $s_x$ and $s_z$ such that
\begin{equation}
e^{-s_x\mathcal{T}_1} e^{-s_z\mathcal{T}_2} X(T; X_0) = X_0.
\label{eqn-2-7}
\end{equation}
We first set up the Newton iteration, although the Newton iteration by
itself is entirely inadequate. Let $\tilde{X}_0$, $s_x$, $s_z$, $T$ be our
initial guess for solving \eqref{eqn-2-8} and let
$Y_0 = e^{-s_x\mathcal{T}_1} e^{-s_z\mathcal{T}_2}X(T; \tilde{X}_0)$.
Then the relative error in the initial guess is given by
\begin{equation}
\norm{Y_0-\tilde{X}_0}/\norm{\tilde{X}_0}.
\label{eqn-2-8}
\end{equation}
 If we assume 
$\tilde{X}_0+\delta X_0$, $s_x+\delta s_x$, $s_z + \delta s_z$, $T+\delta T$
to be the solution of \eqref{eqn-2-8} that is close to the initial
guess and linearize about the initial guess, we get
\begin{equation}
(\delta s_x) (-\mathcal{T}_1 Y_0) + (\delta s_z) (-\mathcal{T}_2Y_0) + (\delta T)f(Y_0)
+ e^{-s_x\mathcal{T}_1}e^{-s_z\mathcal{T}_2}\frac{\partial X(T;\tilde{X}_0)}{\partial \tilde{X}_0}\,
\delta X_0
= \delta X_0 + (\tilde{X}_0-Y_0).
\label{eqn-2-9}
\end{equation}
The number of equations in the linear system \eqref{eqn-2-9} is the same
as the dimension of $\tilde{X}_0$ or $\delta X_0$. Three more equations are
necessary to have as many equations as unknowns. Those equations are
\begin{align}
\delta X_0^\ast (\mathcal{T}_1 \tilde{X}_0) &= 0 \nonumber\\
\delta X_0^\ast (\mathcal{T}_2 \tilde{X}_0) &= 0 \nonumber\\
\delta X_0^\ast f(\tilde{X}_0) &= 0,
\label{eqn-2-10}
\end{align}
where $\delta X_0^\ast$ denotes the transpose of $\delta X_0$.
Since the Navier-Stokes equations for plane Couette flow are unchanged
by shifts in the $x$ and $z$ directions, shifting $\tilde{X}_0$ in the $x$ or
$z$ directions will only shift $Y_0$ in the $x$ and $z$ directions.
The first two equations in \eqref{eqn-2-10} above require that the
correction $\delta X_0$ to $\tilde{X}_0$ must not have components that shift
$\tilde{X}_0$ infinitesimally in the $x$ or $z$ directions. If the correction
$\delta X_0$ slightly advances $\tilde{X}_0$ along the flow induced by
$\dot{X} = f(X)$, the corrected velocity field will remain on the same
orbit. The third equation in \eqref{eqn-2-10} requires that the
correction $\delta X_0$ must have no component along $f(\tilde{X}_0)$. The
equations \eqref{eqn-2-9} and
\eqref{eqn-2-10} together constitute the Newton system.

It is convenient to rewrite the Newton system as $M
\sigma = \rho$, where $\sigma = (\delta X_0; \delta s_x; \delta s_z;
\delta T)$ and $\rho = (Y_0-\tilde{X}_0; 0; 0; 0)$ are both column vectors.
The structure of $M$ follows from \eqref{eqn-2-9} and \eqref{eqn-2-10}.

As evident from \eqref{eqn-2-9}, the application of $M$ to $\sigma$
requires the computation of $\frac{\partial X(T; \tilde{X}_0)}{\partial \tilde{X}_0}
\delta X_0$. This directional derivative can be computed using 
differences to about $7$ digits of accuracy, which is entirely 
adequate.

\begin{center}
\textit{2.3 Finding the locally constrained optimal hook step}
\end{center}
The dimension of the linearized system $M$ can exceed $3\times 10^5$
for periodic solutions such as $P_3$, and it is impractical to
form the Newton system explicitly. The Newton step $\sigma$ can
be found, however, using a Krylov subspace method such as
GMRES which is described in \citep{TrefethenBau} for example.

Often periodic solutions and steady states are computed within a
bifurcation-continuation scenario as in \citep{SNGS}.  We are not in
such a scenario here and the Newton step by itself never leads to
convergence.  The widely used expedient of damping the Newton step is
also ineffective.

The locally constrained optimal hook step is based on the idea
that given a radius $r$ within which we trust the linearization,
the best step $\sigma$ is obtained by minimizing $\norm{M\sigma-\rho}$
subject to the constraint $\norm{\sigma} \leq r$ \citep{DS}. 
The radius of the trust region is varied from step to step by comparing
the actual reduction in error following the step with the reduction in error
predicted by the linearization \citep{DS}.

We show how to compute the locally constrained optimal hook step
within a Krylov subspace. Let $Q_d$ and $Q_{d+1}$ be matrices with $d$
and $d+1$ orthonormal columns obtained by applying GMRES with the
starting vector $\rho$.  The columns of these matrices are orthonormal
bases for Krylov subspaces of dimensions $d$ and $d+1$.  Before trying
to find the locally constrained optimal hook step, we make sure that
$d$ is large enough to permit a solution of the Newton equation
$M\sigma = \rho$ with small relative residual error given
by $\norm{M\sigma-\rho}/\norm{\rho}$.  Let $H_{d+1,d}$ be the
upper Hessenberg matrix that satisfies $M Q_d = Q_{d+1} H_{d+1,
d}$. We minimize $\norm{H_{d+1,d} \sigma_d - Q_{d+1}^\ast \rho}$
subject to the constraint $\norm{\sigma_d} \leq r$. The solution of
this minimization problem for $\sigma_d$ using the singular value
decomposition of $H_{d+1,d}$ is feasible because $d$ is typically a
number smaller than $30$. Once $\sigma_d$ is found, the locally
constrained optimal hookstep is given by $Q_d\sigma_d$.

\begin{center}
\textit{2.4 Computing traveling wave solutions}
\end{center}
The method for computing relative periodic solutions can be modified
to compute traveling wave solutions. Only minimal modifications are
necessary. Instead of allowing the period $T$ to vary from iteration
to iteration, the first modification is to fix $T$ at a value that is
definitely smaller than the period of any periodic or relative
periodic solution. The second modification is to drop the last of the
three equations in \eqref{eqn-2-10} to get a square system of equations
for $\delta X_0$, $\delta s_x$, and $\delta s_z$. The speed of the
traveling wave in the streamwise and spanwise directions will be
given by $s_x/T$ and $s_z/T$, respectively.

The basis of this approach for computing traveling wave solutions is the
observation that the initial velocity field of a traveling wave solution
is a fixed point of the time $T$ map of the flow, if the final velocity
field is shifted by appropriate amounts $s_x$ and $s_z$ in the 
streamwise and spanwise directions. The shifts depend upon the
wave speeds as indicated in the previous paragraph.

In some cases, symmetries of the velocity field may imply that the
traveling wave must actually be a steady solution
\citep{Waleffe5}. For computing traveling waves in general
we solve for $s_x$ and $s_z$ to determine the wave speeds. But if it
is known in advance that the wave speeds are zero, we set $s_x=s_z=0$
and drop the first two equations of
\eqref{eqn-2-10}.  We moved  some of Waleffe's traveling wave solutions
\citep{Waleffe5}, whose data are posted publicly, from a $32\times
34\times 32$ grid to finer grids, with $(2L,M,2N) = (48, 73, 48)$ or
better, and then refined the solutions to better accuracy. During the
refinement the period $T$ was fixed at $1$ or $5$ or $10$, with larger
values of $T$ implying faster convergence of the GMRES iteration. The
refined solutions too were invariant under the shift-reflection and
shift-rotation symmetries.

\section{Verifiability of computed relative periodic solutions}
\begin{figure}
\begin{center}
\includegraphics[height=2in, width=2.5in]{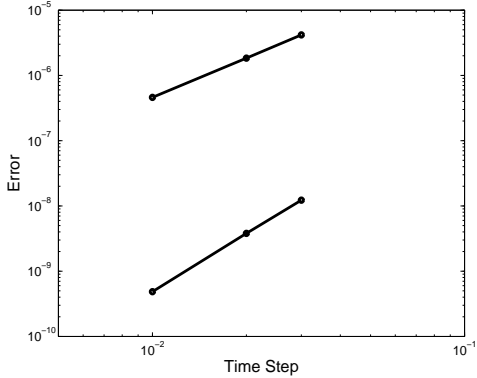}
\end{center}
\caption[xyz]{The plot above shows the dependence of the global
error, obtained by integrating $P_1$ for one full period, on the time step
for numerical integrators of $2$nd and $3$rd order. The $2$nd
order integrator used was Crank-Nicolson-Adam-Bashforth and 
the third order method was the $(4,3,3)$ method of \citep{ARS}.}
\label{fig-2}
\end{figure}

Although the solutions reported in Table \ref{table-1} are all
linearly unstable, they can all be verified using a good DNS code
for channel flow as will be shown in this section. We also explain how
the error estimates reported in the last column of Table \ref{table-1}
were derived.

The characteristic multipliers $\lambda_{max}$ reported in Table
\ref{table-1} are all greater than $1$ but less than $100$ in
magnitude. For characteristic multipliers in that range, the errors
due to time discretization can be made negligible. To see that,
consider the initial value problem $\dot{x} = f(x)$, $x(0) = x_0$. The
global error after time $T$ is defined as $\norm{\tilde{x}(T; x_0) -
x(T; x_0)}$, where $\tilde{x}(T;x_0)$ is the numerical approximation
to the solution $x(T;x_0)$ at time $T$. For an integrator of order $r$ the
global error is asymptotically equal to $E(T)h^r$ in the limit of
small $h$. Indeed, explicit formulas can be obtained for the function
$E(T)$
\citep{Viswanath0}. Those formulas indicate that $E(T)$ will 
increase with $\lambda_{max}$ for the solutions listed in Table \ref{table-1}.
However, as indicated by the asymptotic formula and as demonstrated in
Figure \ref{fig-2}, the global error can be made quite small by
taking a small time step. For all the six solutions in Table \ref{table-1},
we carried out computations similar to the one shown in Figure \ref{fig-2}
and chose a time step small enough to make the time discretization errors
irrelevant. The final computations were all carried out using a 
$3$rd order implicit-explicit Runge-Kutta method developed by
\cite{ARS}.

The argument in the previous paragraph would have been invalid if some
characteristic multipliers were too large in magnitude. For instance,
if $\lambda_{max}>10^{15}$, even the rounding errors would be
amplified to an $O(1)$ magnitude over a single cycle. In such
situations, one has to use multiple shooting. 

Thus if the system $\dot{X} = f(X)$ is obtained by spatially
discretizing the Navier-Stokes equation \eqref{eqn-2-1} and we
compute a relative periodic solution of that system with a small
enough time step, how close the computed solution is to a true
periodic solution of the Navier-Stokes equation is entirely
determined by the spatial discretization error. 

\begin{figure}
\begin{center}
\includegraphics[height=1.8in,width=1.8in]{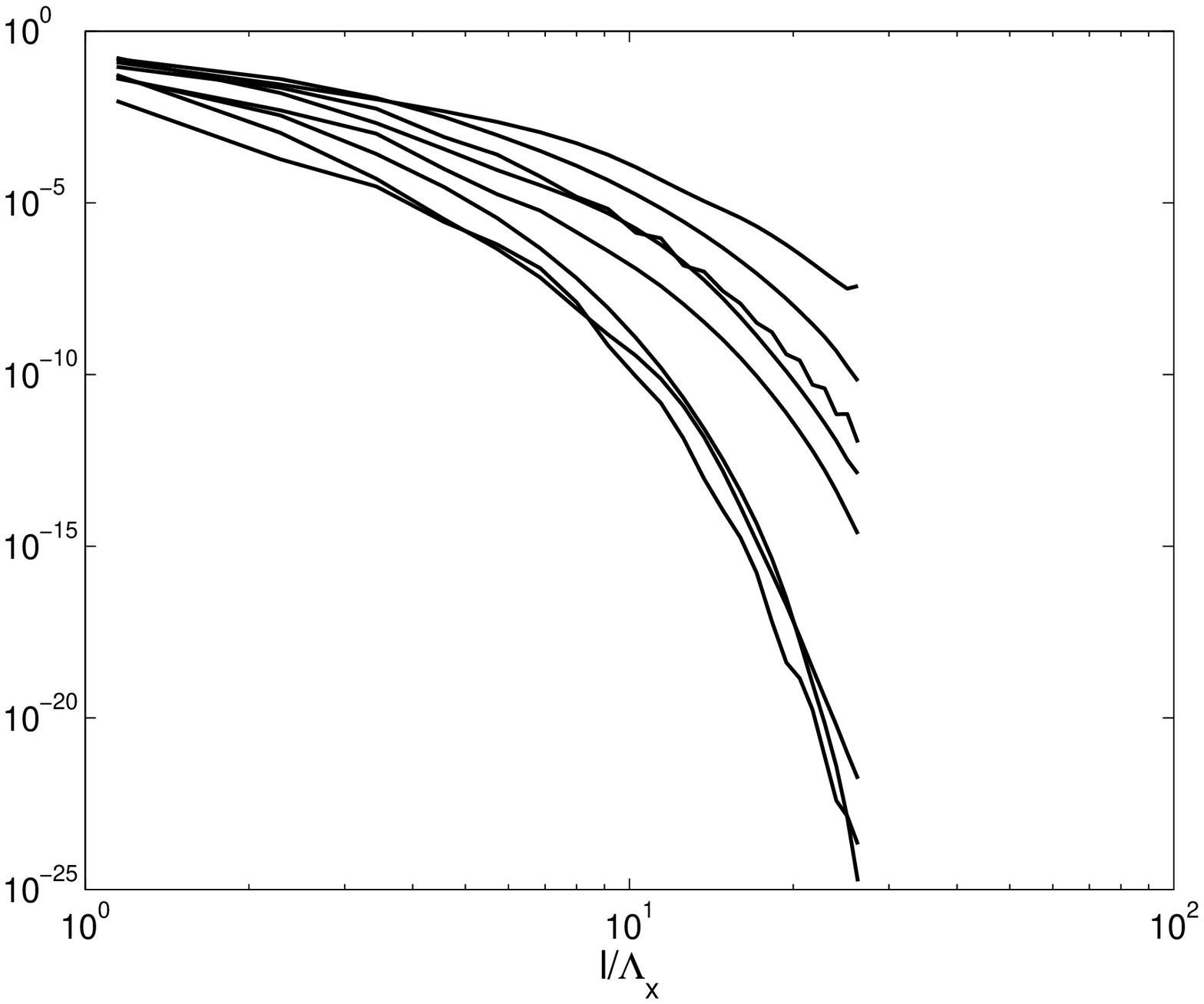}
\includegraphics[height=1.8in,width=1.8in]{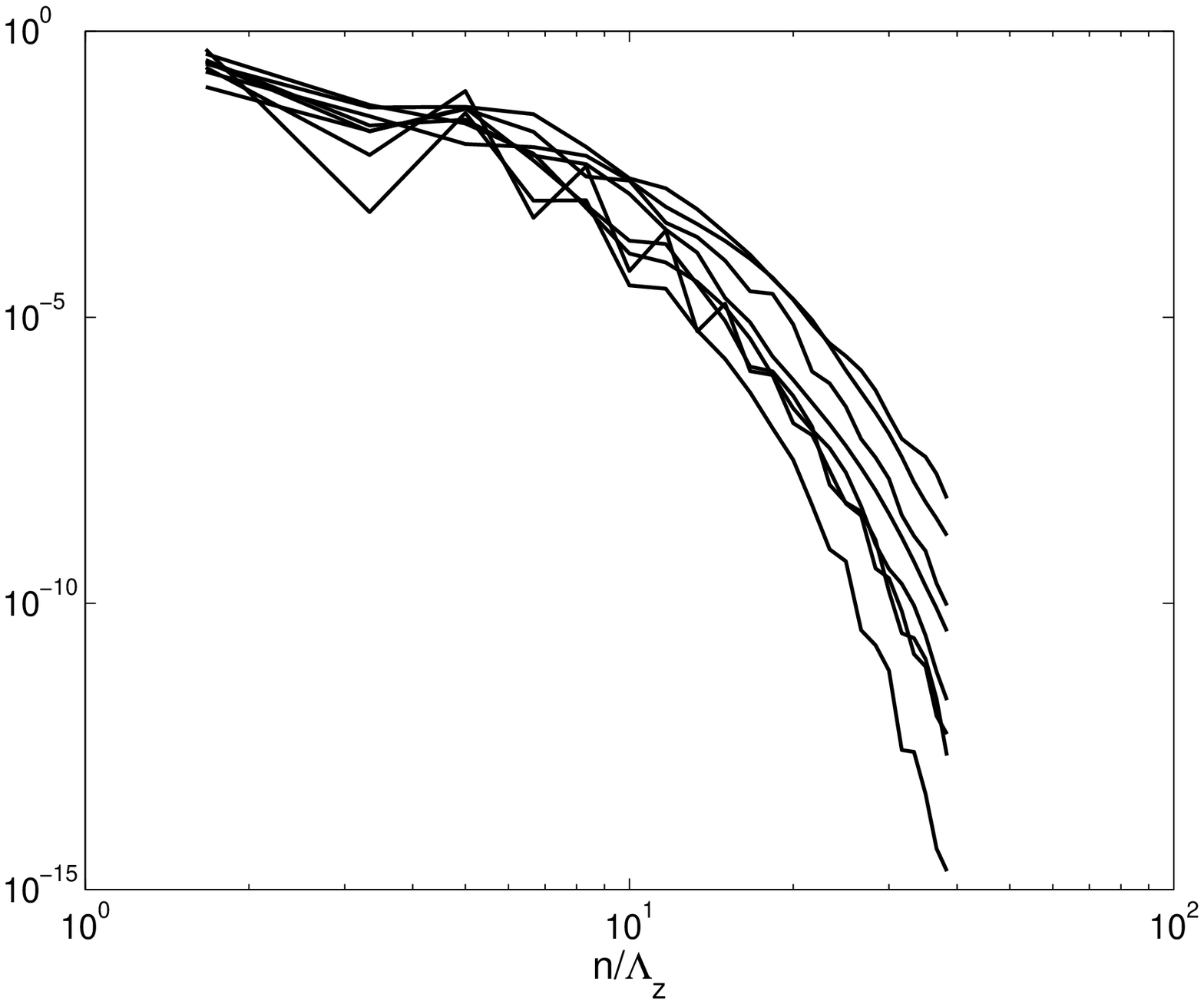}
\includegraphics[height=1.8in,width=1.8in]{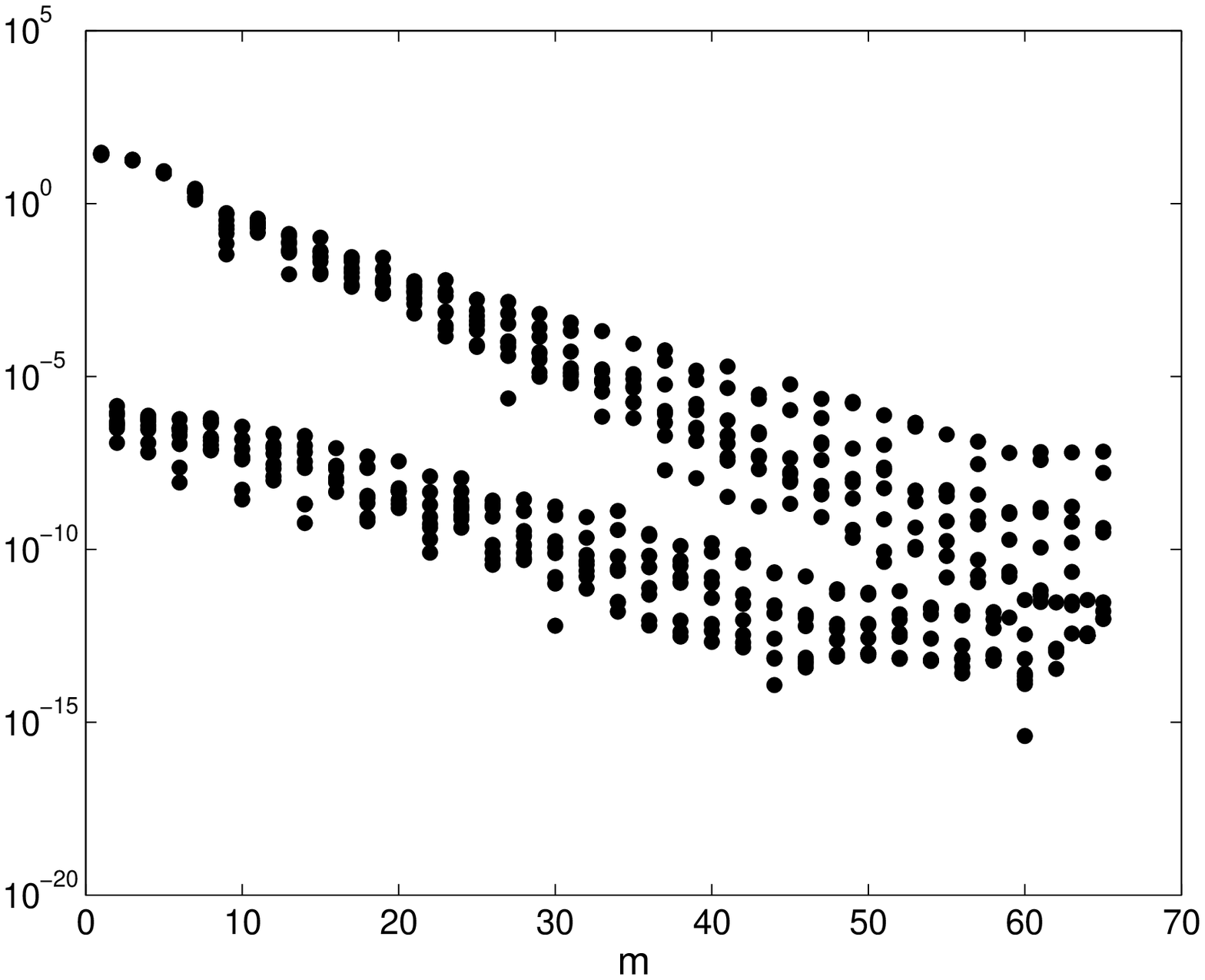}
\end{center}
\caption[xyz]{The first two plots show the variation of the energy 
with streamwise and spanwise wavenumbers, respectively, for the
relative periodic solution $P_6$. The energies are computed using the
slice $y=0$ and at eight equally spaced instants along $P_6$'s
period. The third plot shows the magnitude of the energy in
wall-normal Chebyshev mode against the Chebyshev mode at eight
equally spaced instants along $P_6$'s period.}
\label{fig-3}
\end{figure}

We estimate the spatial discretization error in two ways. The first
way is to graph energy against streamwise wavenumber, spanwise wavenumber,
and wall-normal Chebyshev mode as shown in Figure \ref{fig-3}.
The first two plots in that figure do not show the energy that
corresponds to the wavenumber $0$. Because the Chebyshev polynomials 
are not orthogonal with respect to the Lebesgue measure, the 
decomposition of the energy into Chebyshev modes is necessarily
somewhat arbitrary. We defined
\begin{equation*}
E(y) = \int_0^{2\pi \Lambda_z}\int_0^{2\pi \Lambda_x} 
u(x,y,z)^2 + v(x,y,z)^2 + w(x,y,z)^2 \: dx\, dy,
\end{equation*}
and expressed $E(y)$ as a linear combination of Chebyshev polynomials
to get the third plot in Figure \ref{fig-3}.
More specifically, if $E(y) = \sum_{m=0}^M c_m T_m(y)$, where
$T_m(y)$ are Chebyshev polnomials, the third plot in Figure \ref{fig-3}
plots $\abs{c_m}$ against $m$.
 Estimates for the
discretization error are obtained by taking the square root of the
fraction of the energy in the highest mode. Thus we will have 
streamwise, spanwise, and wall-normal estimates at each point
along the relative periodic solution. The worst of
these estimates corresponds to the direction that is less
well-resolved than the other two, and the instant along the solution
where the velocity field is hardest to resolve.

Another possibly more reliable way to estimate the spatial
discretization error is to take the initial state of the computed
relative periodic solution and then move it to a much finer grid. The
initial data is integrated on this finer grid using a sufficiently
small time step for one full period.  The final state is then shifted
using the shifts $s_x$ and $s_z$.  The 
quantity
\begin{equation*}
\frac{\norm{\text{shifted final state}-\text{initial state}}}
{\norm{\text{initial state}}}
\end{equation*}
is taken as an estimate of the spatial discretization
error.

In all six cases reported in Table \ref{table-1}, these two methods
gave comparable estimates for the spatial discretization error.  All
the errors reported in Table \ref{table-1} were obtained using the
second method and a finer grid with $(2L, M, 2N) = (64, 90, 64)$.  As
our discussion of time discretization error makes it clear, all the
error estimates can be verified using a good DNS code with a
sufficiently small time step. Gibson used data about $P_1$ that had a
relative error of $10^{-5}$ and verified that error estimate using his
{\it Channelflow} code \citep{Gibson}.

\section{Relative periodic solutions and the bursting phenomenon}

The most striking thing about the bursting phenomenon in experiments
is its recurrent nature. As \cite{AS} state ``The study of
boundary-layer turbulence in the last thirty years has clearly
demonstrated that the chaotic behavior referred to as turbulence has
a systematic organization. Most researchers share a common belief that
a cyclic bursting phenomenon is the predominant mode of turbulence
production.'' In this section, we demonstrate the relevance of the
periodic and relative periodic solutions listed in Table \ref{table-1}
to the bursting phenomenon.

\begin{center}
\textit{4.1 Position of relative periodic solutions in phase space}
\end{center}
\begin{figure}
\begin{center}
\includegraphics[height=2.5in,width=2.5in]{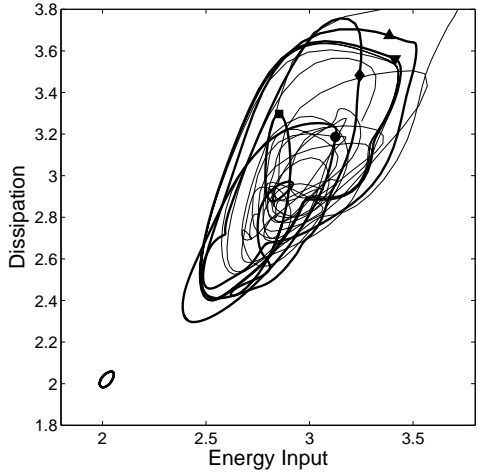}
\end{center}
\caption[xyz]{Energy dissipation ($D$) and energy input ($I$) are
defined by \eqref{eqn-2-3} and \eqref{eqn-2-4}. The periodic orbit in
the lower left corner is $P_1$. The other orbits correspond to the
solutions $P_2$ through $P_6$ with the correspondence given by the
markers. The marker for each periodic orbit is listed in Table \ref{table-1}.
A random turbulent trajectory is shown in the background.}
\label{fig-4}
\end{figure}

A good preliminary idea of the relative location of solutions in phase
space can be formed by projecting the orbits to the $D$-$I$ plane
\citep{KK}. The normalization used for $D$ and $I$ in \eqref{eqn-2-3}
and \eqref{eqn-2-4} is such that the laminar solution of plane Couette
flow is located at $(1,1)$. We immediately see from Figure \ref{fig-4}
that $P_1$ is much closer to the laminar solution than the other
solutions. As mentioned earlier, $P_1$ is not a bursting solution, while
the others are.

Greater energy dissipation is connected with steeper gradients in the
flow.  It is therefore tempting to conclude that orbits that travel
farther into the upper-right corner of Figure \ref{fig-4} are harder
to resolve.  That conclusion is true only approximately. Plots of
energy against wavenumber or Chebyshev mode look very similar to
Figure \ref{fig-3} for $P_2$, $P_3$, $P_4$, and $P_5$, but not for
$P_1$. The most striking aspect of the plots for $P_6$ in Figure
\ref{fig-3} lies in the plot of energy against streamwise
wavenumber. While at one instant the energy falls by more than a
factor of $10^{20}$ over the range of streamwise wavenumbers
represented in the computational grid, it falls by only a factor of
about $10^{8}$ at another instant. This striking change in energy
distribution is directly related to the breakdown of streaks and is
observed in $P_2$ through $P_6$ but not in $P_1$.

The existence of periodic and relative periodic solutions is related
to two general principles in dynamics, namely the Poincar\'{e}
recurrence theorem and the closing lemma \citep{KH}.  The recurrent
nature of the bursting process was observed in direct numerical
simulation of plane Couette turbulence and motivated the derivation of
the self-sustaining process \citep{HKW, Waleffe2}.  It must be pointed
out, however, that the right notion of recurrence in plane Couette
flow is not that of periodicity but that of relative periodicity. This
is because if an initial velocity field for plane Couette flow is
shifted in either the streamwise or the spanwise direction, the final
velocity fields after a certain interval of time must be equal to each
other after exactly the same shifts. From the point of view of the
dynamics, velocity fields that are related by streamwise and spanwise
shifts are equivalent.

The same general principles that suggest the existence of periodic and
relative periodic solutions that correspond to bursting, also suggest
that there will be infinitely many of them. Indeed, these solutions
probably give the best or only means to uncover the ``systematic organization''
found within turbulent boundary layers that Acarlar and Smith refer to
in the passage quoted at the beginning of this section. This possibility
will be discussed at length in the concluding discussion section.

\begin{center}
\textit{4.2 Near-wall statistics}
\end{center}

\begin{figure}
\begin{center}
\includegraphics[height=2.8in, width=2.8in]{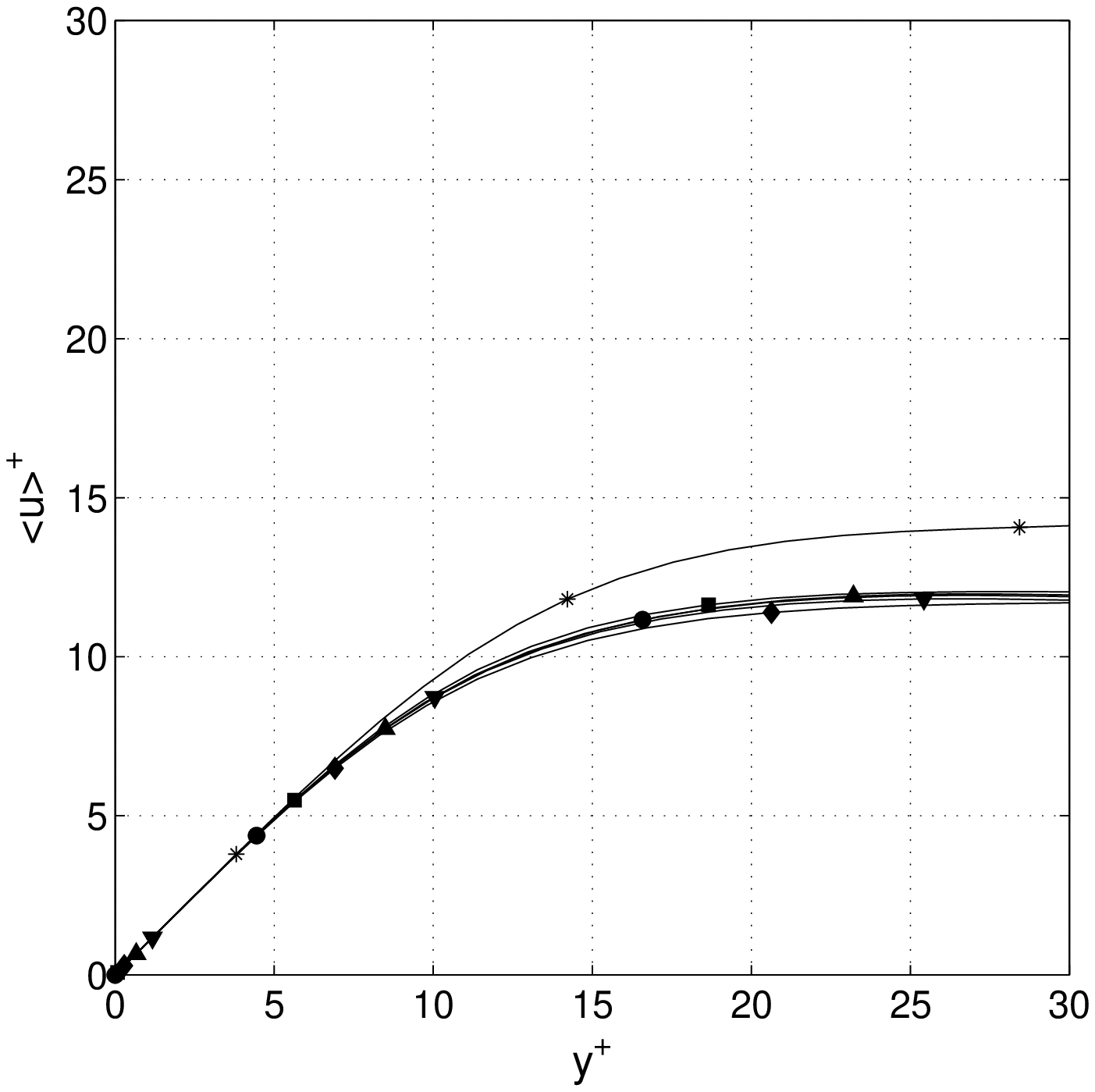}
\includegraphics[height=2.8in, width=2.8in]{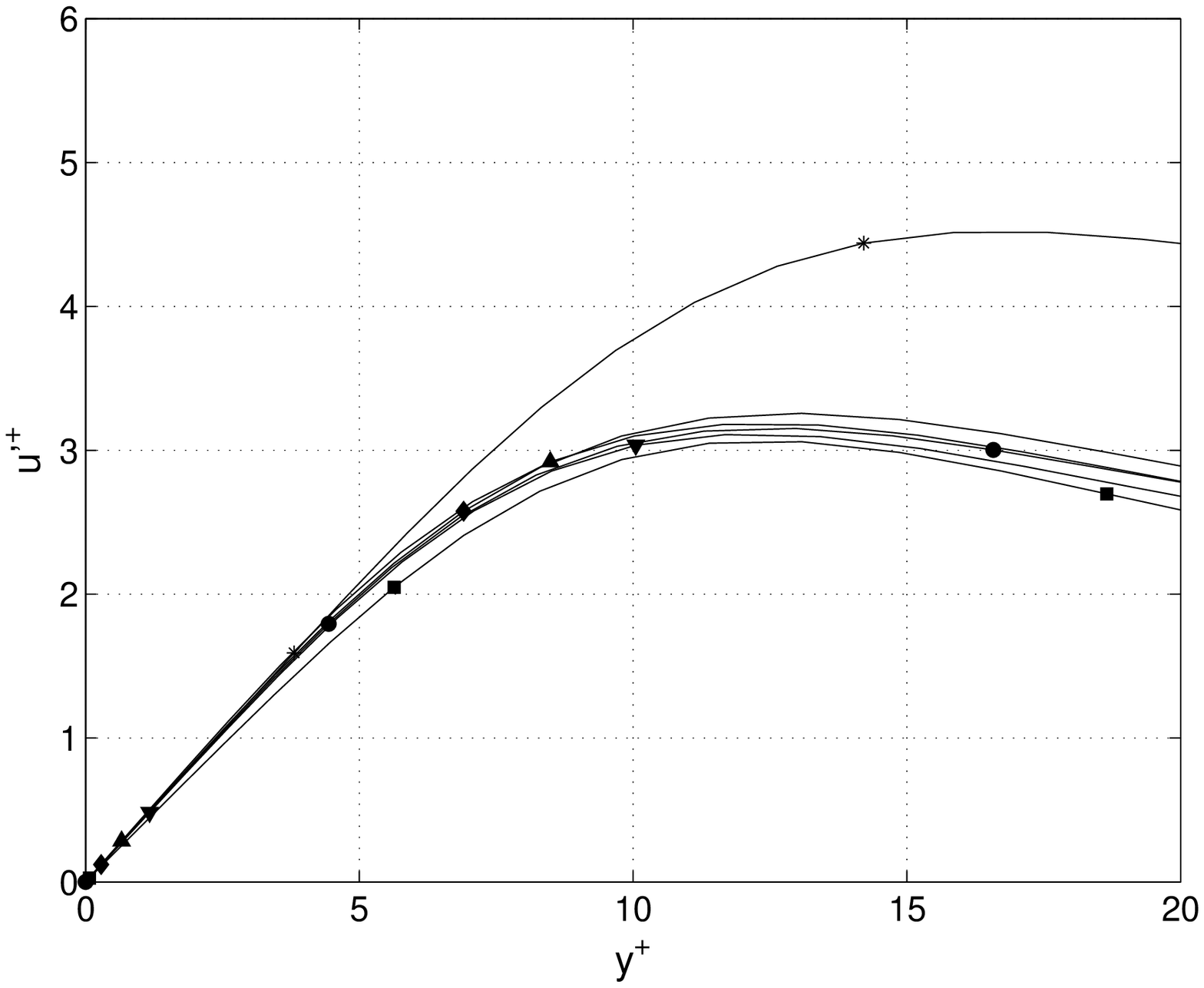}\\
\includegraphics[height=2.8in, width=2.8in]{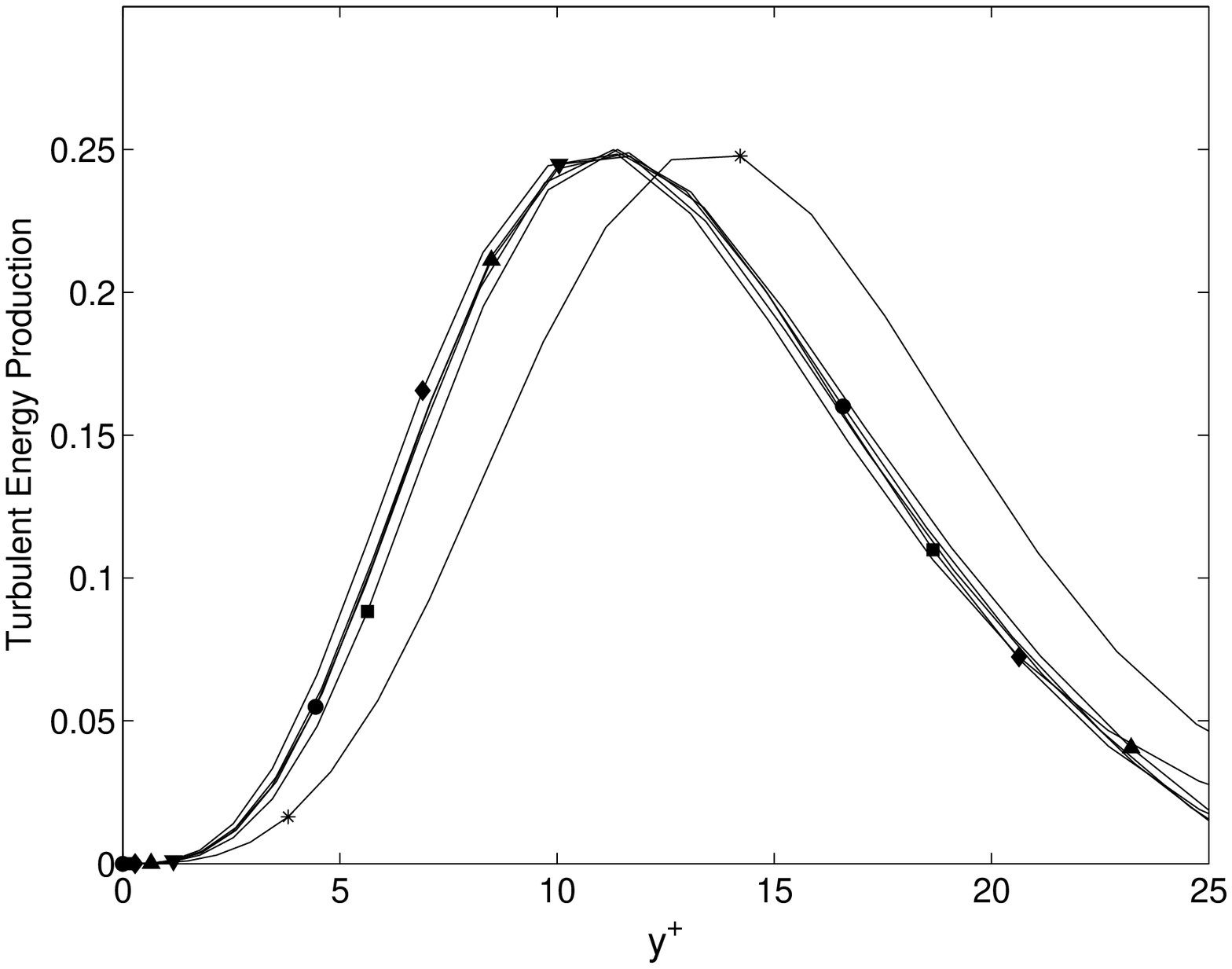}
\end{center}
\caption[xyz]{The first of the top two plots shows
the dependence of the mean streamwise
velocity $<\!\!u\!\!>^+$ in wall units upon the distance $y^+$ from the upper
wall, also in wall units. The second plot shows the dependence of the
turbulent intensity as given by the root mean square streamwise
velocity $u'^+$ upon the distance from the wall. The bottom plot
shows the dependence of turbulent energy production upon the distance
from the wall. The correspondence to the solutions $P_1$ through 
$P_6$ listed in  Table \ref{table-1} is shown  using markers.}
\label{fig-5}
\end{figure}

Turbulent boundary layers can be divided into a viscous sublayer,
a buffer layer, and a logarithmic boundary layer \citep{MY}. In the
viscous sublayer, which is approximately $5$ wall units thick,
the mean streamwise velocity nearly equals the distance from the wall in
wall units. The buffer layer extends from $5$ wall units to approximately
$25$ or $30$ wall units, and the logarithmic boundary layer 
begins after that. The viscous stresses dominate in the viscous sublayer,
while the Reynolds stresses dominate in the logarithmic boundary layer.
Both stresses are significant in the buffer layer.

The first plot in Figure \ref{fig-5} shows the dependence of the mean
streamwise velocity upon the distance from the wall. The averages are
computed over one single period for each periodic solution in all
plots in Figure \ref{fig-5}. The law of the wall is reproduced
correctly in the viscous sublayer, but the logarithmic boundary layer
is not fully developed. The distance between the two moving walls in
plane Couette flow in only about $70$ wall units or so for the
solutions listed in Table \ref{table-1}, which is one reason the
logarithmic boundary layer is not fully developed. Another related
reason is that the frictional Reynolds numbers listed in Table
\ref{table-1} are too low for the logarithmic boundary layer to be
fully formed. A dynamical investigation of the interaction of
structures away from the buffer region with structures in the buffer
region can be found in \citep{ITx}.

The most important features of bursting are in the buffer
layer and these features are reproduced correctly by $P_2$
through $P_6$ as shown by the other two plots in Figure \ref{fig-5}.
The top-right plot graphs the turbulent intensity against the
distance from the wall. The graphs for $P_2$ through $P_6$ all
have the right shape and the turbulent intensity peaks between
$10$ and $20$ wall units for each of those solutions. However,
the peak is slightly elevated as compared with the
theory and experiments recorded in Figure 27 of
\cite{MY} or with the ``corrected'' experiment and
computation recorded in Figure 7 of \citep{KMM}. This elevation of
the peak is a low Reynolds number effect. The peak of the turbulence
intensity compares well with Figure 3 of \citep{JKSNS}.

The energy balance equation, which is obtained by applying the method
of averaging to the Navier-Stokes equation, is important both in
theory and in practice \citep{MY}. Physical interpretations can be
associated with the terms of that equation, and we will look at the
so-called turbulent energy production term.  This term equals $$ -
<\!\!u^\ast v^\ast\!\!> \frac{\partial <\!\!u\!\!>}{\partial y},$$
where $u^\ast = u - <\!\!u\!\!>$ and $v^\ast = v - <\!\!v\!\!>$ are
the fluctuating components of the streamwise and wall-normal
velocities and $<\!\!u\!\!>$ is the mean streamwise velocity.  In 
Figure \ref{fig-5}, the turbulent energy production is expressed in wall
units, although experimentalists do not always use wall units for expressing
turbulent energy production
\citep{KRSR}.

Turbulent energy production has a sharp peak in the buffer region for
each of $P_2$ through $P_6$, as shown by the bottom plot in 
Figure \ref{fig-5}. Turbulent energy production can be readily measured
in experiments and its sharp peak in the buffer region has 
intrigued experimentalists for a long time \citep{KRSR}. The significance
of the bursting phenomenon is in part  because of its connection
to turbulent energy production. We observe from the bottom plot of
Figure \ref{fig-5} that for $P_1$, which is not a bursting solution,
turbulent energy production attains its maximum value farther away from
the wall.

\begin{center}
\textit{4.3 Break-up and advection of coherent structures}
\end{center}

\begin{figure}
\begin{center}
\includegraphics[height=2.1in, width=2.8in]{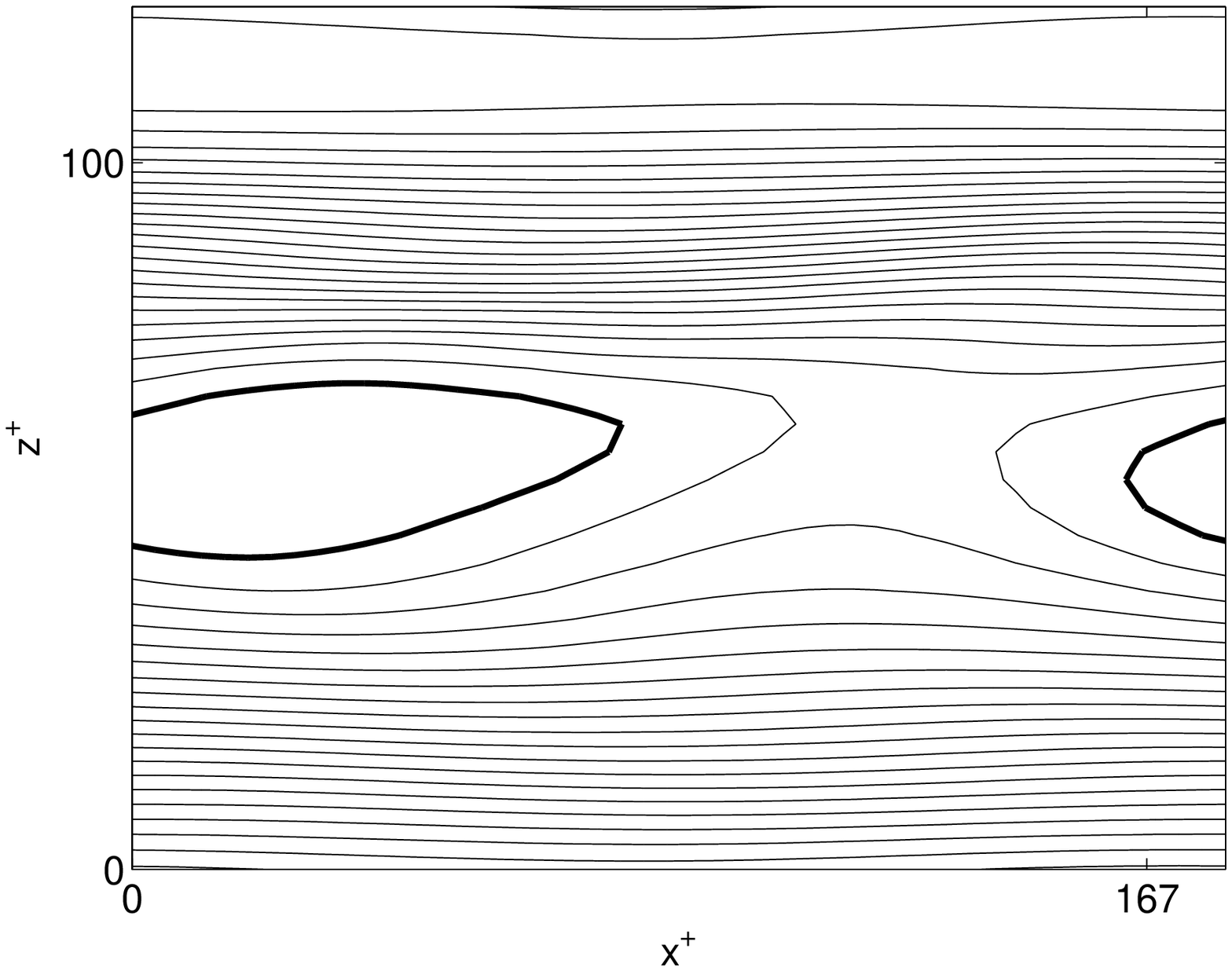}
\includegraphics[height=2.1in, width=2.8in]{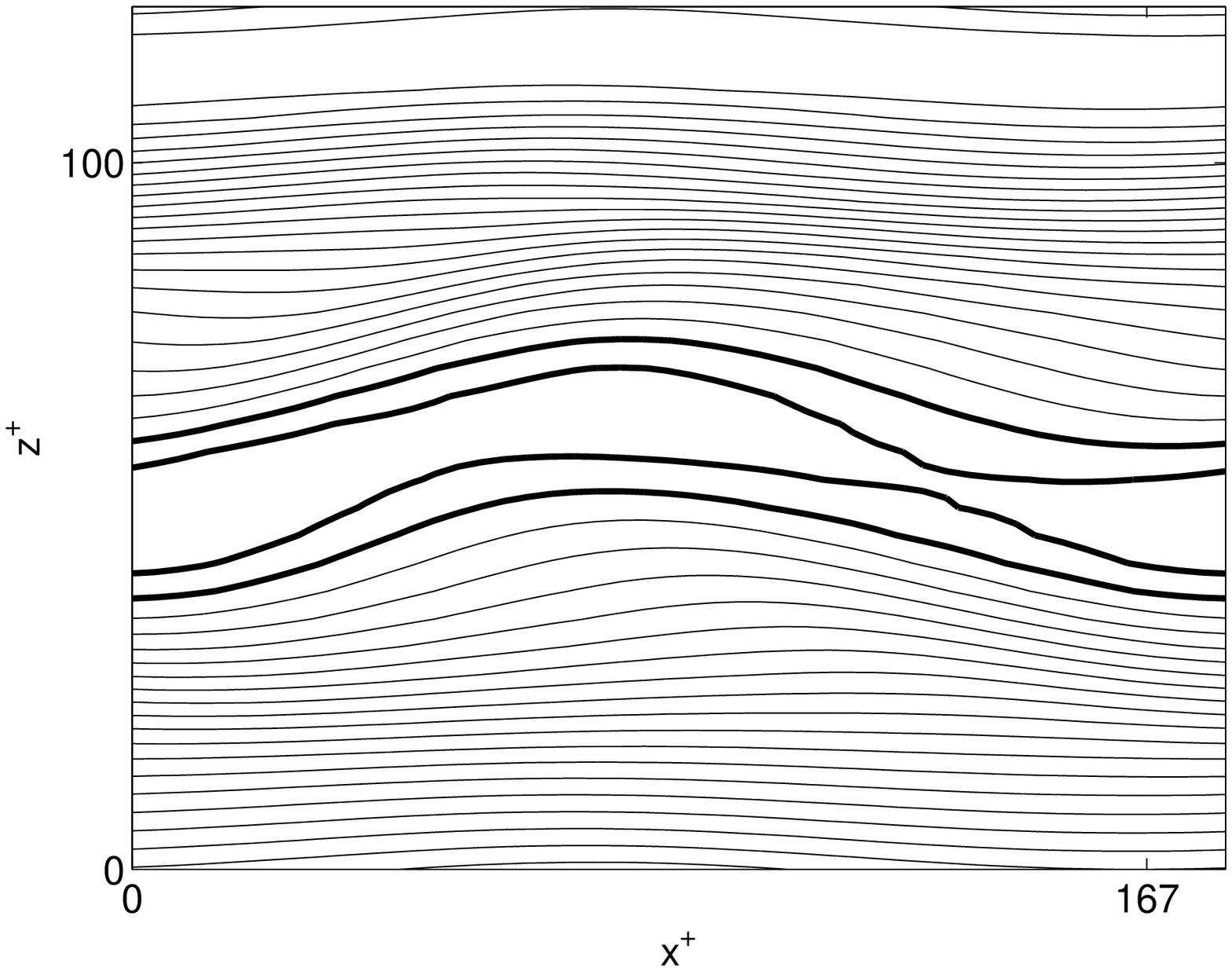}\\
\includegraphics[height=2.1in, width=2.8in]{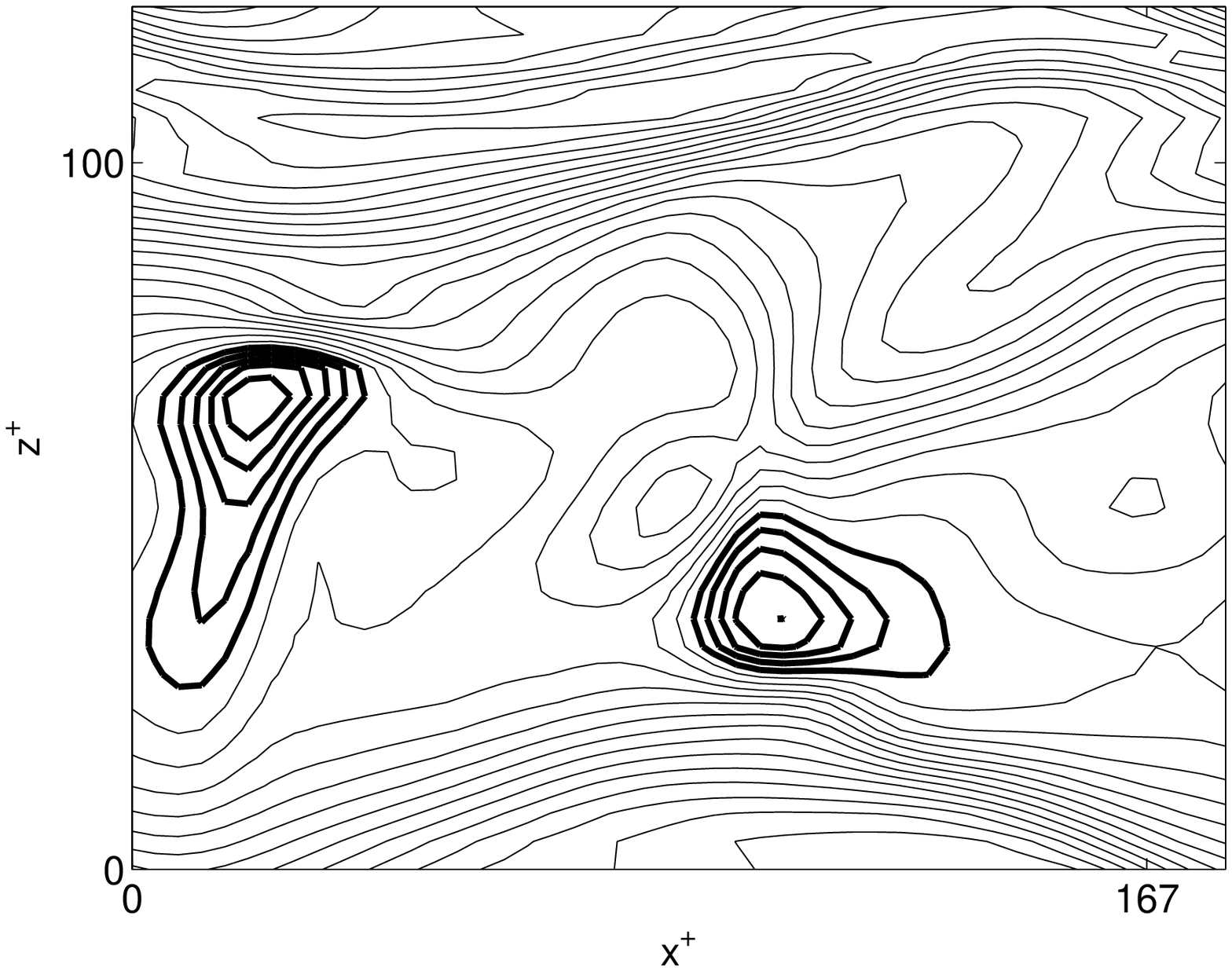}
\includegraphics[height=2.1in, width=2.8in]{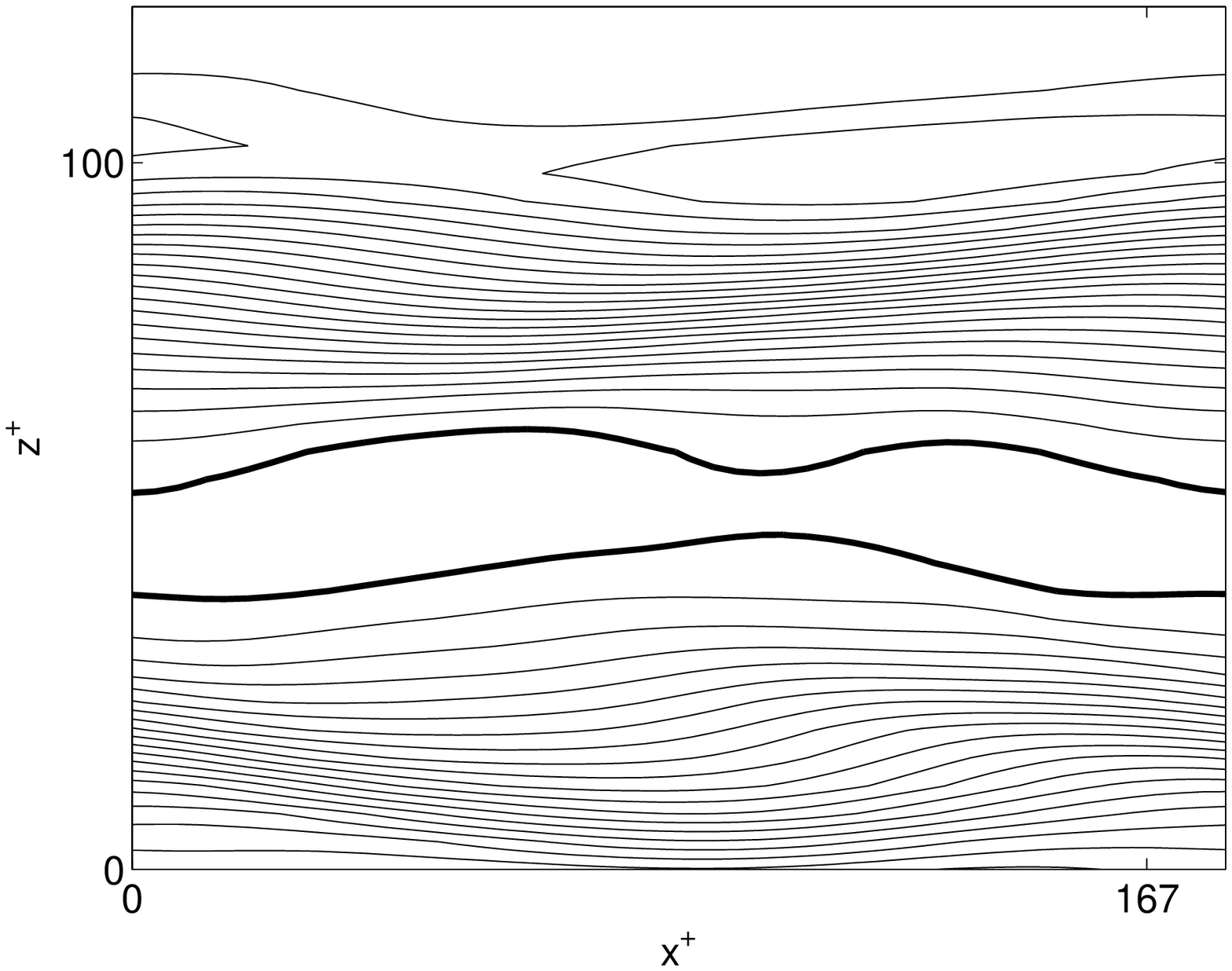}
\end{center}
\caption[xyz]{Isolines of the streamwise velocity in the
$z$ vs. \!\!$x$ plane at $y^{+}\approx 10$ for the relative periodic
solution $P_4$. The four plots correspond to $t=0, T/4, T/2, 3T/4$,
where $T$ is the period of $P_4$.
The plots are in clockwise order, beginning at the top left.
The velocity fields were shifted in
such a way that a plot at $t=T$ would coincide exactly with the plot
for $t=0$. The isolines are thin if the streamwise velocity is
positive, and thick if it is negative. Each of the four plots has $24$
isolines equispaced between minimum values which are $-0.5, -1.4,
-2.3, -0.5$ (in wall units) and maximum values which are $9.4, 9.4,
7.5, 8.3$.}
\label{fig-6}
\end{figure}

\begin{figure}
\begin{center}
\includegraphics[height=2.8in, width=2.8in]{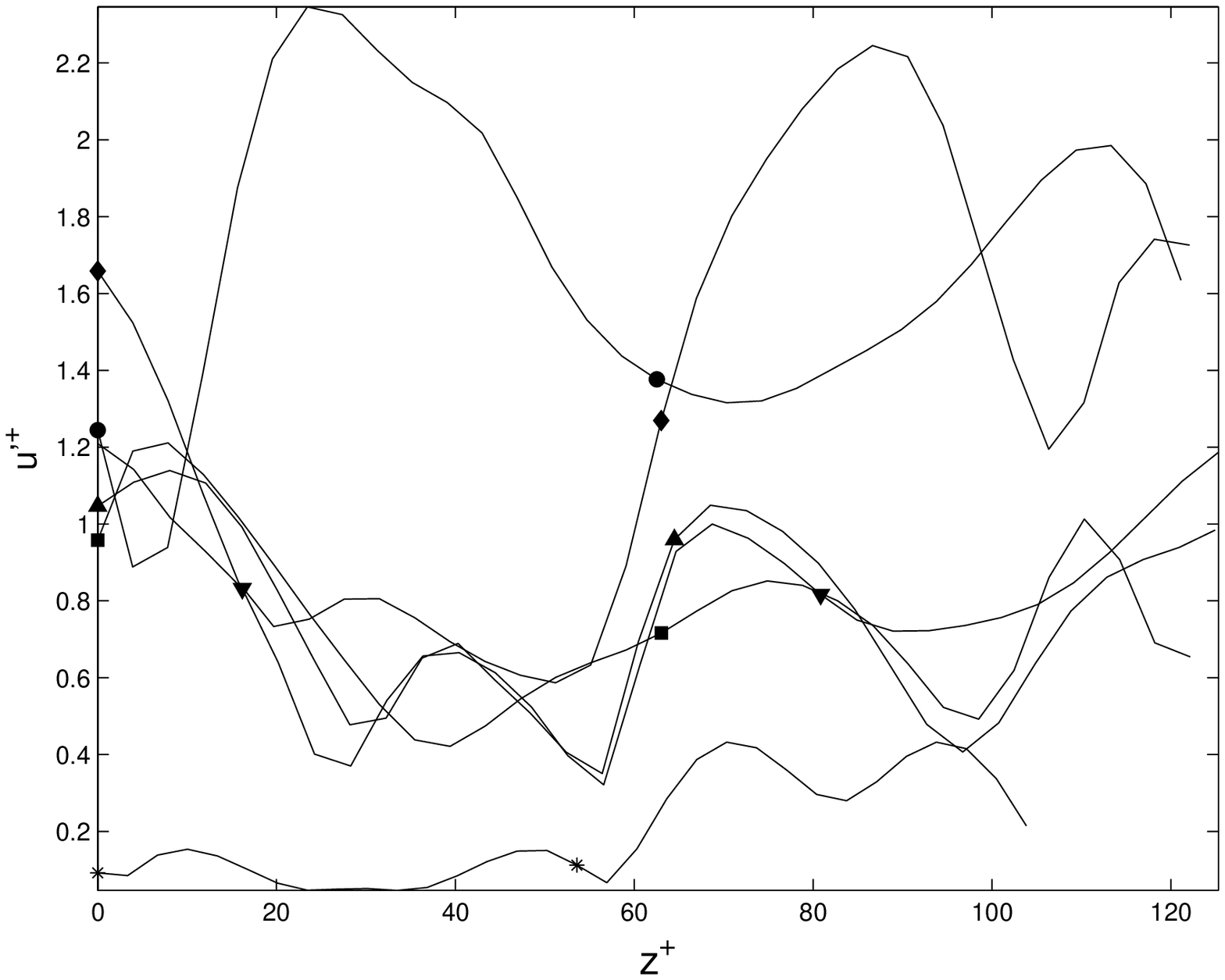}
\end{center}
\caption[xyz]{The plot above shows the variation of the root
mean square value of the fluctuating part of streamwise velocity,
which is denote by $u'^+$, in the spanwise direction. The rms
value is computed at $y^+ \approx 10$ and $x=\pi\Lambda_x$. }
\label{fig-7}
\end{figure}

Streaks are the most prominent coherent structures observed in
turbulent boundary layers. In streaky velocity fields, the streamwise
velocity is only weakly dependent on the streamwise coordinate but
varies much more strongly in the spanwise direction. In contour plots such as
those in Figure \ref{fig-6}, streakiness shows up as isolines that are
nearly parallel to the $x$ axis. It is clear from that figure that 
streaks at $t=T/2$ break-up at $t=3T/4$ and then re-form. The plot
at $t=0$ corresponds to the initial velocity field of $P_4$.

It is difficult to measure the velocity field as a whole in experiments,
and therefore experimental visualizations of streaks rely on hydrogen
bubbles introduced into the flow by platinum wires and other techniques
\citep{KRSR, SM}. Sometimes streaks are detected using pointwise
measurements \citep{KTS}. Figure \ref{fig-7} shows the dependence
of the fluctuations of the streamwise velocity on the spanwise
direction. The rms velocity $u'^+$ shown in the top-right plot
of Figure \ref{fig-5} was obtained by averaging over a period
and by averaging spatially in the spanwise and streamwise directions.
But in Figure \ref{fig-7}, only the time average was used for computing
the rms velocity.

The possibility that the break-up of streaks observed experimentally
could be an artifact of flow visualization techniques has been
considered by \cite{JKSNS}. It has been suggested that the advection
of permanent objects could be partly responsible for experimental
observations. Our computation of relative periodic solutions ($P_2$
through $P_5$) shows that temporal periodicity and the advection of
permanent objects are not mutually exclusive possibilities. While it
is generally implicitly assumed that the advection of coherent structures could
only be in the streamwise direction, our computation of $P_2$ and
$P_4$ shows that the advection could be in the spanwise direction as
well. This could be significant, as will be seen shortly.

The spanwise variation in the strength of the fluctuations of the streamwise
velocity shown in Figure \ref{fig-7} for $P_1$ through $P_6$ is
reminiscent of experimental data, such as Figure 2 of \citep{KTS}
for instance. This spanwise variation is not very pronounced for $P_1$,
which is not a bursting solution, but is very pronounced for $P_2$
and $P_4$. Those are the only two solutions in Table \ref{table-1} which
have a spanwise drift. We are led to conclude that spanwise advection
of coherent structures could be a significant source of the observed
spanwise variation  of $u'^+$.

\begin{center}
\textit{4.4 Discrete symmetries of plane Couette flow}
\end{center}
\begin{figure}
\begin{center}
\includegraphics[height=2.1in,width=2.8in]{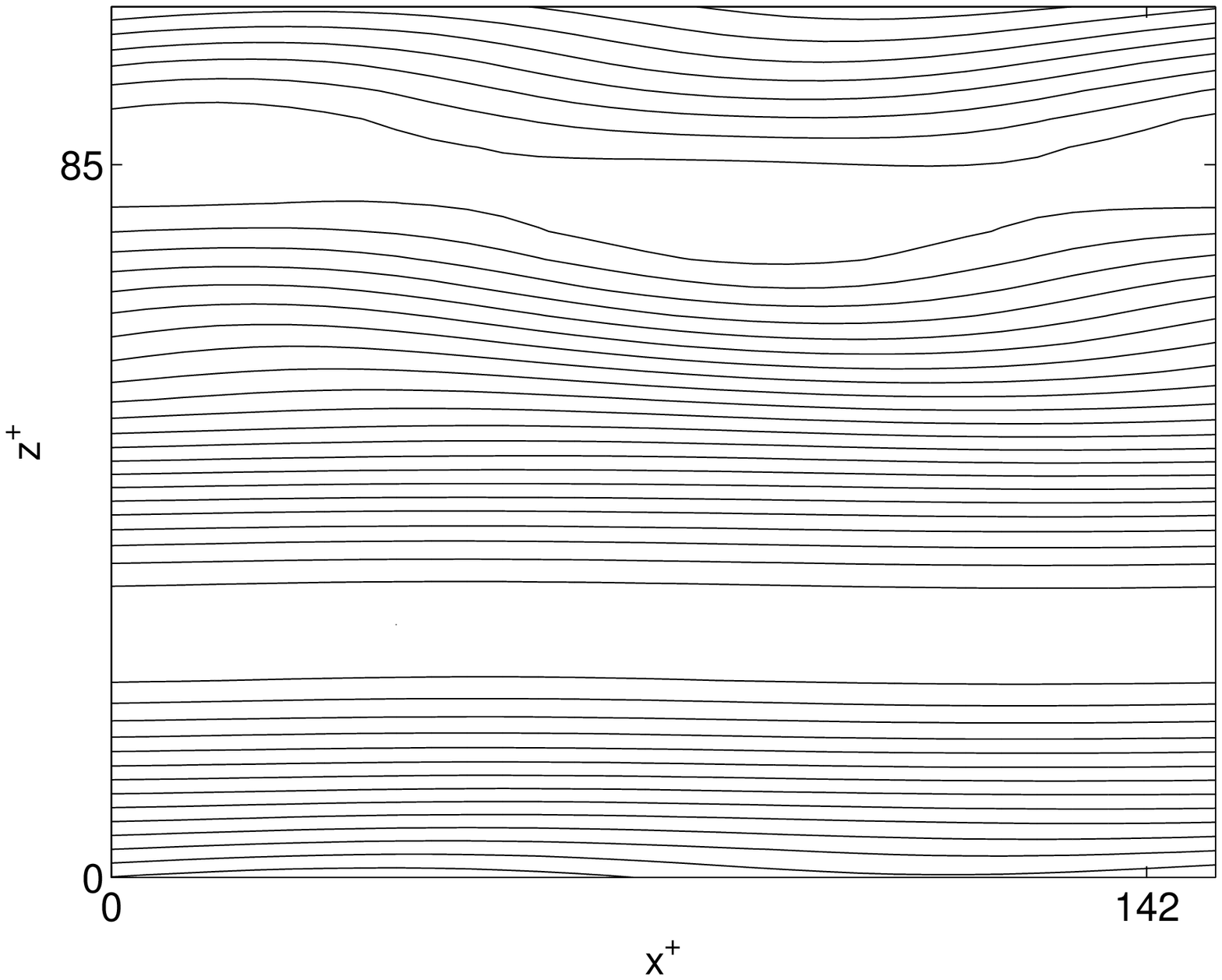} 
\includegraphics[height=2.1in,width=2.8in]{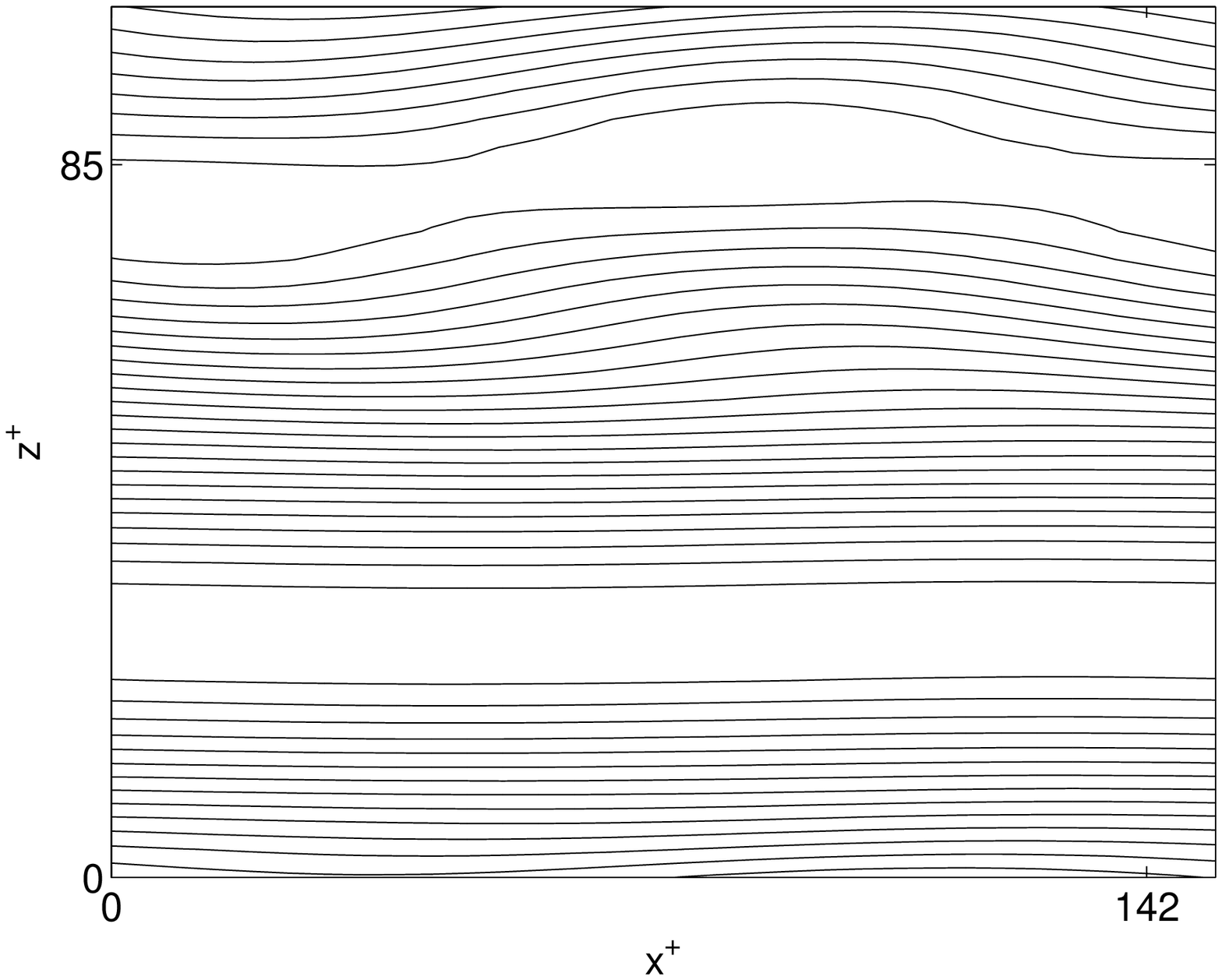} 
\end{center}
\caption[xyz]{The two plots above show slices through the streamwise
velocity field of $P_1$ at $t=0$ and $t=T/2$, respectively. The
slices were taken at $y^+\approx 9$. As explained in the text,
the plots show that $P_1$ does not have the shift-reflection symmetry.
Isolines of $u^+$ are drawn at $24$ equispaced values between a maximum
of $11.6$ and a minimum of $1.6$. The maximum occurs in the wide gap between
isolines in the lower part of either plot, and the minimum occurs in the
gap in the upper part.
}
\label{fig-8}
\end{figure}

The Navier-Stokes equation for plane Couette flow has two
discrete symmetries \citep{Waleffe5}. The shift-reflection
transformation of the velocity field is given by
\begin{equation*}
\begin{pmatrix}
u\\v\\w
\end{pmatrix}
\Biggl(x+\pi \Lambda_x, y, -z\Biggr),
\end{equation*}
and the shift-rotation transformation of the velocity field is
given by
\begin{equation*}
\begin{pmatrix}
-u\\-v\\w
\end{pmatrix}
\Biggl(-x+\pi \Lambda_x, -y, z + \pi \Lambda_z\Biggr).
\end{equation*}
Plane Couette flow is unchanged under both these transformations.
Thus if a single velocity field along a trajectory of plane Couette
flow satisfies either symmetry, all points along the trajectory must
have the same symmetries. However, velocity fields that lie on
the stable and unstable manifolds of symmetric periodic or relative
periodic solutions need not be symmetric.

Figure \ref{fig-8} shows that $P_1$ does not have the shift-reflection
symmetry. Both the plots would look very different if they were
flipped upside-down and shifted forward by half the width of the plot
in the $x$ direction. We have verified that $P_1$ does not have the
shift-rotation symmetry either. A close inspection of Figure
\ref{fig-8} reveals that the second plot can be obtained by shifting
the first plot in the $x$ direction. In fact, $P_1$ is a relative
periodic solution. During half the period listed in Table
\ref{table-1}, the initial velocity field of $P_1$ shifts by $s_x/\pi\Lambda_x = 0.5$ and $s_z=0$.

The averaged velocity field of a long turbulent trajectory of plane
Couette flow satisfies the discrete symmetries to a very rough
approximation. However, there is no reason to believe that such an
average has any dynamical significance. The average could just be a
transient state and it may not even lie on the asymptotic set of the
flow. 

The six solutions listed in Table \ref{table-1} satisfy
neither the shift-reflection symmetry nor the shift-rotation symmetry.
This implies that the shift-reflection and shift-rotation transformations
(which commute with each other) can be applied to each  solution
in Table \ref{table-1} to obtain $18$ more solutions.

\begin{center}
\textit{4.5 Unstable manifolds of the solutions $P_i$}
\end{center}
\begin{figure}
\begin{center}
\includegraphics[height=2.8in, width=2.8in]{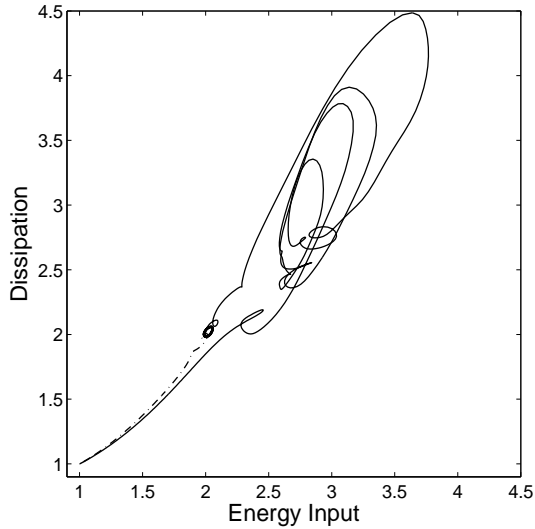}
\end{center}
\caption[xyz]{The plot above shows the projection of the unstable
manifold of $P_1$ to the energy dissipation vs.\!\! energy input plane.}
\label{fig-9}
\end{figure}

In their study of the bursting phenomenon 
in turbulent Poiseuille flow, \cite{IT} found that the break-up
and re-formation of coherent structures is well approximated by
the unstable manifold of a traveling wave solution. Among the solutions
we have computed, $P_1$ most resembles the traveling wave solution of
\citep{IT}. Motivated by that resemblence, we asked if the solutions
$P_2$ through $P_6$ could be related to the unstable manifold of $P_1$.

 We perturbed $P_1$ along the single unstable direction that
corresponds to that solution. A perturbation in one sense begins to
head towards the laminar solution right away and falls into the
laminar solution.  A perturbation with the opposite sign heads into
the turbulent region, as shown in Figure \ref{fig-9} but
relaminarizes eventually.  In contrast, we found that perturbations
along the unstable manifolds of the bursting periodic solutions $P_2$
through $P_6$ do not relaminarize. Thus there appears to be no
relationship between the unstable manifold of $P_1$ and the other
solutions.

In addition, unlike $P_1$ the bursting solutions have more than one
unstable direction.  For instance, the Arnoldi iteration shows that
$P_2$ has at least $11$ unstable directions. These facts suggest that
the unstable manifold of $P_1$ cannot by itself explain the
bursting phenomenon in the plane Couette scenario considered in this
paper.

\section{Discussion}

The approach to the bursting phenomenon, and to turbulence in fluid
flows more generally, that this paper advocates is quite simple.  As
the incompressible Navier-Stokes equation is an excellent physical
model, our approach is simply to compute solutions of that
differential equation. This approach should not be surprising in
itself because the main reason for deriving differential equations is
to understand their solutions. However, well-resolved spatial
discretizations of turbulent phenomena require more than a hundred
thousand degrees of freedom and computing solutions with so many
degrees of freedom is nontrivial. We have shown that problems
associated with largeness of the number of degrees of freedom can be
overcome. Indeed, our computations of periodic and relative periodic
solutions used as many as twenty times the number of degrees of
freedom in any earlier computation of periodic solutions.

Having decided to compute solutions, one must decide what type of
solutions to look for within turbulent flows. Here the basic 
principles of dynamics are of help. The recurrent nature of the
bursting phenomenon has been noted by experimentalists, as we
pointed out earlier. Recurrence, which is the tendency of certain
dynamical systems to revisit points close to their initial state
in phase space after extended excursions, has long been a central
and unifying idea in dynamics. The relevance of recurrence to 
long term dynamics is captured in a most general way by the Poincar\'{e}
recurrence theorem \citep{KH}. In its early days, this theorem gave
rise to troubling questions about the foundations of statistical
mechanics. The number of degrees of freedom needed for resolving
low Reynolds number turbulence is quite large, yet far smaller
than the number of degrees of freedom typical of statistical mechanics.
Another significant difference is that low Reynolds number turbulence
is governed by equations that are not Hamiltonian and in which
viscosity damps high wavenumbers. Therefore there is reason to think
that a point of view based on recurrences will be useful for
understanding turbulence.

A combination of local instability, which is  a well-known feature
of turbulent flows, and boundedness in phase space naturally suggests
the existence of periodic motions. The closing lemma is a principle
that suggests that dynamics can be understood quite generally in terms
of periodic solutions \citep{KH}. It has been proved in a few
restricted settings and serves as a beacon.  While considering the
closing lemma, it is worth noting again that the right notion of
recurrence depends upon invariance properties of the underlying
differential equation. If the differential equation is unchanged by a
continuous group of transformations, then it is appropriate to look
for relative periodic solutions.

In some well-understood settings, it is possible to prove the
existence of infinitely many periodic solutions \citep{Moser}. Even
though turbulent phenomena are not known to fall under any of these
settings, there is reason to think that there are infinitely many
periodic and relative periodic solutions embedded within turbulent
flows.  The lack of complete theoretical results should not be an
impediment to computation, since the major purpose of computation is
to render tractable problems that are beyond the reach of theory.  An
advantage of computing many of these solutions is that they can help
us understand the dynamics as a whole in terms of accurately computed
periodic solutions. In addition, these
periodic solutions could serve as a basis to understand turbulent statistics
using the periodic orbit theory
\citep{CAMTV}.

It is natural to expect different solutions of the same differential
equation to bear a relationship to one another. In the case of linear
differential equations, the principle of linear superposition gives
this relationship. For chaotic nonlinear systems, the relationship
between solutions is much more complicated and intriguing. In several
instances, the precise nature of the relationship is given by symbolic
dynamics \citep{Moser}. Algorithms for
computing nonlinear systems can be based upon symbolic dynamics
\citep{Viswanath1, Viswanath2}. Those algorithms made it possible to
compute the fractal structure of the well-known Lorenz
attractor. Although the fractal structure of the Lorenz attractor was
deduced by \cite{Lorenz}, its computation became possible only with
algorithms based on symbolic dynamics. The Lorenz equations were
derived to illuminate the essentially deterministic nature of
turbulence and they have been completely successful in that
respect. Computations of the strange attractor of the Lorenz equations
can likewise serve as  models for computations of turbulence.

The Lorenz computations based on symbolic dynamics illustrate the
advantanges of periodic solutions over steady solutions in
understanding the dynamics as a whole. It is possible to accurately
compute periodic solutions that are as close as machine precision
permits to random points on the Lorenz attractor, and thus obtain a
very good understanding of the dynamics \citep{Viswanath1}. Such a
precise understanding cannot be obtained using steady solutions alone.
To understand important aspects of turbulent boundary layers such as
the turbulent energy production in the buffer layer, it is likewise
necessary to go beyond steady solutions and traveling waves and compute
periodic and relative periodic solutions.

It would be premature to suggest that the various solutions embedded within
turbulent flows can be described using symbolic dynamics. Yet, it would
be very surprising if these solutions did not bear a relationship to
one another. Discovering such a relationship would be a major advance
in our understanding of turbulent flows.

This paper has asserted the existence of six solutions of turbulent
plane Couette flow. Error estimates for these solutions were
supplied in Table
\ref{table-1}.  We argued in Section 3 that these solutions, along
with their error estimates, can be verified using a good code for direct
numerical simulation of channel flows. Such quantitative
reproducibility is a step forward in the area
of  turbulence computation.

\section{Acknowledgments}
The author thanks P. Cvitanovi\'{c} and F. Waleffe for many helpful
discussions; J.F. Gibson for checking some of the solutions using his
{\it Channelflow} code; and G. Kawahara and S. Toh for
clarifications.   This work was supported by the NSF
grant DMS-0407110 and by a research fellowship from the Sloan
Foundation.

Data for solutions reported in this article can be
obtained by contacting the author. 

\bibliography{references}
\bibliographystyle{plainnat}
\end{document}